\documentclass[eqsecnum]{article}

\usepackage{amsfonts}
\usepackage{ulem}
\usepackage{float}
\usepackage{amssymb}
\usepackage{amsmath}
\usepackage{mathrsfs} 
\usepackage[all]{xy}
\usepackage[dvips]{color}
\usepackage[dvips]{graphicx}

\newcommand{\omits}[1]{}

\definecolor{dyellow}{rgb}{1.,0.8,.0}
\definecolor{myblue}{rgb}{.1,.1,.7}
\definecolor{dcyan}{rgb}{.0,.6,.6}
\definecolor{dmagenta}{rgb}{0.6,0.0,0.6}
\definecolor{brown}{rgb}{0.6,0.2,0.}
\definecolor{darkblue}{rgb}{.0,.0,0.5}
\definecolor{darkred}{rgb}{0.75,0.0,0.0}
\definecolor{orange}{rgb}{1.,.6,.0}
\definecolor{dorange}{rgb}{0.8,.4,.0}
\definecolor{darkgreen}{rgb}{0.0,0.6,0.0}
\definecolor{purple}{rgb}{.4,.0,.4}
\definecolor{grey}{rgb}{0.7,0.7,0.7}

\newcommand{\hide}[1]{}
\newcommand{\delete}[1]{}
\newcommand{\distribute}[3]{}	

\newcommand{\od}{\mathrm{d}}

\title{\textbf{Hamiltonian Analysis of 4-dimensional Spacetime in Bondi-like Coordinates}}

\author{Chao-Guang Huang\thanks{Email: \text{huangcg@mail.ihep.ac.cn}},
and Shi-Bei Kong\thanks{Email: \text{shibeikong@ecut.edu.cn}},
\\
Theoretical Physics Division, Institute of High Energy Physics,
\\
Chinese Academy of Sciences, Beijing 100049, China,\\
and University of Chinese Academy of Sciences, Beijing 100049, China}

\begin{document}

\maketitle

\begin{abstract}

We discuss the Hamiltonian formulation of gravity in 4-dimensional spacetime under Bondi-like coordinates $\{v,r,x^{a},a=2,3\}$.
In Bondi-like coordinates, the 3-dimensional hypersurface is a null hypersurface and the evolution direction is the advanced time $v$.
The internal symmetry group SO(1,3) of the 4-dimensional spacetime is decomposed into SO(1,1), SO(2), and T$^{\pm}$(2),
whose Lie algebra $\mathfrak{so}$(1,3) is decomposed into $\mathfrak{so}(1,1),\mathfrak{so}(2),$t$^{\pm}(2)$ correspondingly.
The SO(1,1) symmetry is very obvious in this kind of decomposition, which is very useful in $\mathfrak{so}(1,1)$ BF theory.
General relativity can be reformulated as the 4-dimensional coframe ($e^{I}_{\mu}$) and connection ($\omega^{IJ}_{\mu}$) dynamics of gravity based
on this kind of decomposition in the Bondi-like coordinate system.
The coframe consists of 2 null 1-forms $e^{-},e^{+}$ and 2 spacelike 1-forms $e^{2},e^{3}$. The Palatini action is used.
The Hamiltonian analysis is conducted by the Dirac's methods.
The consistency analysis of constraints has been done completely.
There are 2 scalar constraints and one 2-dimensional vector constraint.
The torsion-free conditions are acquired from the consistency conditions of the primary constraints about $\pi^{\mu}_{IJ}$.
The consistency conditions of the primary constraints $\pi^0_{IJ}=0$ can be reformulated as Gauss constraints.
The conditions of the Lagrange multipliers have been acquired.
The Poisson brackets among the constraints have been calculated.
There are 46 constraints including 6 first class constraints $\pi^0_{IJ}=0$ and 40 second class constraints.
The local physical degrees of freedom is 2.
The integrability conditions of Lagrange multipliers $n_{0},l_{0}$, and $e^{A}_{0}$ are Ricci identities.
The equations of motion of the canonical variables have also been shown.

\bigskip

\end{abstract}

\tableofcontents

\section{Introduction}

The Hamiltonian analysis plays an extremely important role in the initial-value problem and canonical quantization.  For a gravitational system,
the Hamiltonian analysis depends on two fundamental elements, namely, the foliation of a spacetime and the choice of canonical variables.

The most frequently used foliation is to foliate a spacetime by  a series of 3-dimensional spacelike hypersurfaces along a timelike vector field \cite{ADM}, based on which the initial-value problem is well defined.  An alternative foliation is to foliate the spacetime along two null vector fields \cite{d'IS}, named by 2+2 formalism, based on which the initial-value problem can also be well defined.

In order to understand the gravitational radiation better, a null foliation is proposed \cite{Goldberg1}, which provides a canonical formulation of
a theory on outgoing null hypersurfaces. In a neighborhood of an outgoing beam of wave near the future null infinity in an asymptotical flat spacetime,
the metric can be written
in a Bondi-Sachs coordinate system $\{u,r,x^{a},a=2,3\}$ \cite{Bondi,Sachs},
\begin{equation}
\od s^2 = g_{00}\od u^{2}+2g_{01}\od u \od r+2g_{0a}\od u\od x^{a}+g_{ab}\od x^{a}\od x^{b},
\end{equation}
with $g_{00},g_{01}<0,g_{0a}>0$. The metric has 4 Bondi conditions $g_{11}, g_{12}, g_{13}=0$ and $\mathrm{det}(g_{ab})\sim r^{2}$.
In the system, the retarded time $u$ is a null coordinate.  Each $u$ defines a 3-dimensional null
hypersurface in the 4-dimensional spacetime.  The spatial coordinate $r$ is regarded as the distance from the isolated gravitational source.
For a given $u$, every $r$ defines a 2-dimensional spacelike surface in the 3-dimensional null hypersurface.

For a beam of an outgoing gravitational wave, $u$ always keeps constant in its propagation direction.
So, if one wants to study the propagation properties of a given beam of gravitational wave, the advanced time coordinate $v$ should be used
instead of the retarded time coordinate $u$. In the study of the geometry near an isolated horizon which is a null hypersurface, the advanced time coordinate $v$ should also be used \cite{ABF, AK, Krishnan}.
In these cases, the metric is better written in a Bondi-like coordinate system $\{v,r,x^{a},a=2,3\}$ or $\{x^{0},x^{1},x^{a},a=2,3\}$:
\begin{equation}
\od s^2 = g_{00}\od v^2 + 2g_{01}\od v \od r+2g_{0a} \od v \od x^a +g_{ab}\od x^a\od x^{b} \label{1.2},
\end{equation}
with $g_{00}<0$, $g_{01}$, $g_{0a}>0$.  In the above metric, there are 3 Bondi-like conditions $g_{11}$, $g_{12}$, $g_{13}=0$,
so the metric has only 7 variables rather than 10 variables in the general form of a metric. The 4th coordinate condition is not imposed here.

Unlike the 1+3 spacelike foliation and 2+2 foliation which can be used in the analysis of initial-value problems in whole spacetime,
the 1+3 null foliation can be only used in a finite region of the spacetime where there is no null signal incident in the opposite direction.  In fact, in the study of the propagation of a beam of gravitational wave, the one-way propagating wave and its propagation property are focused on and thus it is supposed that there exist no other null signals.  In the case of an isolated horizon, by definition, it is a null hypersurface without the incident of ingoing signals.

In order to have a better knowledge of the evolution of a geometry, the 3-geometry $h_{ij}$ on a 3-dimensional spacelike hypersurface in ADM formalism \cite{ADM} and the 2-geometry $\gamma_{ab}$ on a 2-dimensional spacelike surface in $2+2$ formalism are chosen as the canonical configuration variables.
In order to make general relativity look like a gauge theory, having polynomial forms, $\mathfrak{su}(2)$-connection on a 3-dimensional
hypersurface is chosen as the canonical configuration variable \cite{Ashtekar}.  The $\mathfrak{su}(2)$-connection is also constructed for $2+2$ formalism \cite{Inverno1,Inverno2,Inverno3} and for $1+3$ null decomposition \cite{Goldberg2,Goldberg3}, and serves as the canonical configuration variable.
The reason of the choice of $\mathfrak{su}(2)$-connection comes from that the Lorentz group can be decomposed as the direct product of two SO(3) subgroups, namely, ${\rm SO}(1,3)={\rm SO}(3)\otimes {\rm SO}(3)$, and the corresponding Lie algebra $\mathfrak{so}(3)$ is isomorphic to $\mathfrak{su}(2)$.

Since the local symmetry SO$(1,d-1$) in a $d$-dimensional spacetime with $d\neq4$ does not have the
similar decomposition, such a kind of connection dynamics cannot be generalized to other dimensional spacetimes.
In order to overcome the difficulty, the $\mathfrak{so}(d)$-connection instead of $\mathfrak{so}(d- 1)$-connection is chosen as basic configuration variable \cite{BTT}.  With the $\mathfrak{so}(d)$-connection, unfortunately, the Lagrangian formalism on a spacetime with Lorentzian signature fails to be constructed though the Hamiltonian formalism can be established \cite{BTT}.  In fact, the local Lorentz group SO(1, $d-1$) in a $d$-dimensional spacetime can always be decomposed as SO$(1, d-1)=$SO(1, 1) $\times$ SO$(d-2)\times $T$^{-}(d-2)\times $T$^{+}(d-2)$, where the latter two cross product $\times$ are Cartesian products of the subgroups \cite{Hall}.  Another problem of the decomposition ${\rm SO}(1,3)={\rm SO}(3)\otimes {\rm SO}(3)$ is
that the SO(1, 1) local symmetry does not appear manifestly. The local SO(1,1) symmetry is very essential
in BF-theory approach to the statistical explanation of black hole entropy \cite{WMZ,BF3,BF4,BF5}. Therefore,
it is worthwhile checking the possibility of choosing  $\mathfrak{so}(1,d-1)=\mathfrak{so}(1,1)\oplus\mathfrak{so}(d-2)\oplus\mathfrak{t}^{-}(d-2)\oplus\mathfrak{t}^{+}(d-2)$-connection as
the canonical configuration variable.

The decomposition of $\mathfrak{so}(1,d-1)=\mathfrak{so}(1,1)\oplus\mathfrak{so}(d-2)\oplus\mathfrak{t}^{-}(d-2)\oplus\mathfrak{t}^{+}(d-2)$ can be easily realised in a coframe consisting of 2 null 1-forms ($e^{-},e^{+}$) and $d-2$ spacelike 1-forms $e^{a}$, which is similar to the Newman-Penrose form
\cite{NP}. It is because the coframe has 4 kinds of local transformations: boost, rotation and 2 kinds of translations,
which leave the metric invariant \cite{BCG}. They belong to 4 subgroups of the Lorentz group SO$(1,d-1)$, namely SO(1, 1), SO$(d-2)$, T$^{-}(d-2)$
and  T$^{+}(d-2)$. In particular, the SO(1, 1) symmetry acts on ($e^{-},e^{+}$) only and the SO($d-2$) symmetry acts on $e^{a}$ only.
In a Bondi-like coordinate system near an isolated horizon or a beam of gravitational wave, the null coframe $e^-$ is chosen to be proportional to d$v$, which makes the SO(1, 1) symmetry more obvious.

In our previous paper \cite{HK}, we have carried out the Hamiltonian analysis of 3-dimensional gravity
in Bondi-like coordinates, based on Dirac's treatments of constrained system \cite{Dirac}.
In the 3-dimensional case, $g_{01}$ is fixed to 1, all the 3 variables $e^{+}_{0}$, $e^{2}_{0}$, $e^{2}_{2}$ of the coframe and the connection components $\omega^{IJ}_\mu$ are treated as configuration variables, the Palatini action is used and the cosmological constant is also included.
The consistency analysis has been carried successfully, torsion-free conditions and Gauss constraints are acquired.
There are only second class constraints. The BTZ spacetime is discussed as a test, which satisfies all the constraints.

The aim of the present paper is to make the Hamiltonian analysis of 4-dimensional gravity in Bondi-like coordinates by the  same method as in Ref. \cite{HK}. For convenience, we make some modifications in the treatment. Different from the treatment in the 3-dimensional case \cite{HK},
$g_{01}$ is not fixed, so the metric is more general and can be applied to more cases. The other differences are that $n_0$, $l_0$ and $e^A_0$ are treated as Lagrange multipliers and that the cosmological constant is not included. The consistency conditions of constraints will require the multipliers $n_{0}$, $l_{0}$ and $e^{A}_{0}$ satisfying certain equations. These equations defines the first derivative of $n_{0}$, $l_{0}$ and $e^{A}_{0}$ with respect to different coordinates and, therefore, the multiplier should satisfy integrability conditions.
Such a situation is not met in Dirac's original literature \cite{Dirac}.
In the new approach, the torsion-free conditions will appear as the consistency conditions of primary constraints containing $\pi^{\mu}_{IJ}$.
In the coframe framework, the Gauss constraints are not independent ones and they will emerge in the consistency conditions of $\pi^0_{IJ}=0$.

The arrangement of the paper is as follows.
In Sec.2, the symmetry decomposition, coframe, connection, action and Poisson brackets are introduced.
In Sec.3, the consistency conditions for the constraints are analysed and the equations of motion are obtained.
As a part of consistency conditions, the integrability conditions of $n_{0}$, $l_{0}$ and $e^{A}_{0}$ are also presented.
In Sec.4, the classifications of constraints are dealt with and the local physical degrees of freedom are discussed. The scalar, vector, and  Gauss constraints in the new approach are also given in this section.
In Sec.5, the summary is made.
In Appendix A and B, 2 identities are proved.
In Appendix C, D and E, the integrability conditions of $n_{0}$, $l_{0}$ and $e^{A}_{0}$ are shown to be equivalent to Ricci identities.
The non-zero Poisson brackets among constraints are listed in Appendix F.

\section{Preliminary}

\subsection{Symmetry Decomposition}

The internal symmetry group of the 4-dimensional spacetime is SO(1,3), and its Lie algebra is $\mathfrak{so}$(1,3).
The generators are denoted as $L_{IJ},I,J=0,1,2,3$, satisfying
\begin{alignat}{1}
[L_{IJ},L_{KL}]=\eta_{IL}L_{JK}+\eta_{JK}L_{IL}-\eta_{IK}L_{JL}-\eta_{JL}L_{JK},\label{2.3}
\end{alignat}
where $\eta_{IJ}=\mathrm{diag}(-1,1,1,1)$ is the Minkowski metric of the local space.

The generators of $\mathfrak{so}$(1,3) can also be redefined as \cite{Hall}
\begin{alignat}{1}
L_{-+}:=L_{01},\quad L_{\pm A}:=\frac{1}{\sqrt{2}}(L_{0A}\pm L_{1A}),\quad L_{AB}:=L_{AB},
\end{alignat}
where $A,B=2,3$. They satisfy
\begin{alignat}{1}
[L_{-+},L_{-A}]=&-L_{-A},\quad [L_{-+},L_{+A}]=L_{+A}, \quad [L_{-+},L_{AB}]=0,\quad [L_{-A},L_{-B}]=0,  \notag \\
[L_{-A},L_{+B}]=&L_{AB}-\delta_{AB}L_{-+}, \quad [L_{-A},L_{BC}]=\delta_{AB}L_{-C}-\delta_{AC}L_{-B}, \quad [L_{+A},L_{+B}]=0,\notag \\
[L_{+A},L_{BC}]=&\delta_{AB}L_{+C}-\delta_{AC}L_{+B},\quad [L_{AB},L_{CD}]=\delta_{AD}L_{BC}+\delta_{BC}L_{AD}-\delta_{AC}L_{BD}-\delta_{BD}L_{AC}.
\end{alignat}
The above equations can also be written together as \eqref{2.3} with $I, J = -, +, 2,3$ and
\begin{alignat}{1}
(\eta_{IJ})=\left(     \begin{array}{cccc}
              0 & -1 & 0& 0\\
             -1 & 0 & 0 &0\\
              0 & 0 & 1 &0\\
               0 & 0 & 0 &1\\
            \end{array}
          \right).\label{eq:internal-metric}
\end{alignat}

\subsection{Coframe}

The spacetime line element can be written in terms of coframe,
\begin{alignat}{8}
\od s^{2}=\eta_{IJ}e^{I}\otimes e^{J}.\label{2.7}
\end{alignat}

Corresponding to our decomposition, the coframe is $\{e^{-},e^{+},e^{A}\}$,
which contains two null 1-forms $e^{-},e^{+}$ (or $n,l$) and 2 spacelike 1-forms $e^{A},A=2,3$.
For any coframe like this, the following 4 kinds of gauge transformations \cite{BCG} leave the line element \eqref{2.7} invariant:
\begin{alignat}{8}
&E^{-}=\frac{e^-}{\alpha}, \quad  E^{+}=\alpha e^+,\quad  E^{A}=e^A,\label{1.10}\\
&E^{-}=e^{-}-c_{A}e^{A}+\frac{1}{2}c_{A}c^{A}e^{+},\quad E^{+}=e^+,\quad  E^{A}=e^A-c^{A}e^{+},\label{1.12}\\
&E^{-}=e^-, \quad E^{+}=e^+-b_{A}e^{A}+\frac{1}{2}b_{A}b^{A}e^{-},\quad  E^{A}=e^A-b^{A}e^{-},\label{1.11}\\
&E^{-}=e^{-},\quad E^{+}=e^+,\quad  E^{A}=e^A\cos\beta-\epsilon_{AB}e^{B}\sin\beta,\label{1.13}
\end{alignat}
which correspond to SO(1,1), T$^{-}$(2), T$^{+}$(2) and SO(2) transformations respectively. Here $\alpha$, $b^{A}$, $c^{A}$ and $\beta$
are gauge parameters, which are arbitrary functions of the coordinates.

\subsection{Connection}

Both $e^{I}$ and $E^{I}$ should satisfy torsion-free conditions
\begin{alignat}{1}
\od e^{I}+\omega^{IJ}\wedge e^{K}\eta_{JK}=0,\quad \od E^{I}+\Omega^{IJ}\wedge E^{K}\eta_{JK}=0.
\end{alignat}
If $e^{I}$ and $E^{I}$ are related by gauge transformations \eqref{1.10}, \eqref{1.11}, \eqref{1.12} and \eqref{1.13},
one can get the relations between $\omega^{IJ}$ and $\Omega^{IJ}$:
\begin{alignat}{6}
\Omega^{-+}=&\omega^{-+}-\od\ln\alpha,\quad\Omega^{-A}=\frac{1}{\alpha}\omega^{-A},\quad \Omega^{+A}=\alpha\omega^{+A},\quad \Omega^{AB}=\omega^{AB};\\
\Omega^{-+}=&\omega^{-+}-\omega^{-A}b_{A},\quad \Omega^{-A}=\omega^{-A},\quad \Omega^{AB}=\omega^{AB}+\omega^{-A}b^{B}-\omega^{-B}b^{A},\notag \\
\Omega^{+A}=&\omega^{+A}+\omega^{-+}b^{A}-\omega^{-B}b_{B}b^{A}+\omega^{AB}b_{B}+\od b^{A}+\frac{1}{2}\omega^{-A}b_{B}b^{B};\\
\Omega^{-+}=&\omega^{-+}+\omega^{+A}c_{A},\quad \Omega^{+A}=\omega^{+A},\quad \Omega^{AB}=\omega^{AB}+\omega^{+A}b^{B}-\omega^{+B}b^{A},\notag \\
\Omega^{-A}=&\omega^{-A}-\omega^{-+}c^{A}-\omega^{+B}c_{B}c^{A}+\omega^{AB}c_{B}+\od c^{A}+\frac{1}{2}\omega^{+A}c_{B}c^{B};\\
\Omega^{-+}=&\omega^{-+},\quad \Omega^{\pm A}=\omega^{\pm A}\cos\beta-\epsilon_{AB}\omega^{\pm B}\sin\beta,\quad \Omega^{AB}=\omega^{AB}+\od\beta.
\end{alignat}

\subsection{Action}

In the following analysis, a special coframe is chosen
\begin{alignat}{1}
n=n_{0}\od v,\quad l=l_0 \od v+\od r,\quad e^A=e^A_0\od v+e^A_a\od x^a,\label{1.14}
\end{alignat}
or written as
\begin{alignat}{1}
e^{-}=e^{-}_{0}\od x^{0},\quad e^{+}=e^{+}_0 \od x^{1}+\od r,\quad e^A=e^A_0\od v+e^A_a\od x^a.
\end{alignat}
The 4 dimensional Palatini action of gravity is
\begin{alignat}{1}
S=&\int_{M}F^{IJ}\wedge\Sigma_{IJ}
=\int_{M}\frac{1}{2}\epsilon_{IJKL}\epsilon^{\mu\nu\rho\sigma}F^{IJ}_{\mu\nu} e^K_\rho e^L_\sigma\od v\od x^{1}\od x^2\od x^3
\nonumber \\
=&\int_{M}(\epsilon_{IJKL}\epsilon^{0jkl}F^{IJ}_{0j} e^K_k e^L_l
+\epsilon_{IJKL}\epsilon^{0ijl}F^{IJ}_{ij}e^K_0 e^L_l)\od^4 x,
\end{alignat}
where
\begin{alignat}{1}
F^{IJ}=\od\omega^{IJ}+\eta_{KL}\omega^{IK}\wedge\omega^{LJ}.
\end{alignat}
So the Lagrangian is
\begin{alignat}{1}
L=\int\epsilon_{IJKL}\epsilon^{0ijk}(F^{IJ}_{0i}e^K_j e^L_k+F^{IJ}_{ij}e^K_0 e^L_k)\od^3x.\label{2.22}
\end{alignat}

In the following analysis, $e^{A}_{a}$ and $\omega^{IJ}_{\mu}$ will be treated as configuration variables,
and their conjugate momenta are denoted as $\pi^{a}_{A}$ and $\pi^{\mu}_{IJ}$ respectively.
$e^{I}_{0}$ will be treated as Lagrange multipliers, so there are 4 corresponding primary constraints:
\begin{alignat}{1}
\epsilon_{IJKL}\epsilon^{0jkl}F^{JK}_{jk}e^L_l=\epsilon_{IJKL}\epsilon^{jkl}F^{JK}_{jk}e^L_l\approx0,
\end{alignat}
where $\epsilon^{0jkl}$ is written as $\epsilon^{jkl}$ for short.
Under coframe \eqref{1.14}, the above 4 constraints can be written as
\begin{alignat}{1}
&\epsilon_{AB}\epsilon^{ab}F^{+A}_{1a}e^{B}_{b}+F^{23}_{23}\approx0,\label{2.9}\\
&\epsilon_{AB}\epsilon^{ab}F^{-A}_{1a}e^{B}_b\approx0,\label{2.10}\\
&F^{-A}_{23}+\epsilon^{ab}e^{A}_aF^{-+}_{1b}\approx0, \label{2.11}
\end{alignat}
corresponding to $n_{0}$, $l_{0}$ and $e^{A}_{0}$, respectively.

\subsection{Poisson Bracket}

The Poisson bracket of 2 quantities $f(v,x)$ and $g(v,y)$ at the same time $v$ is defined as
\begin{alignat}{1}
\{f(v,x),g(v,y)\}=&\int[\frac{\delta f(v,x)}{\delta e^{A}_{a}(v,z)}\frac{\delta g(v,y)}{\delta\pi^{a}_{A}(v,z)}
+\frac{1}{2}\frac{\delta f(v,x)}{\delta\omega^{IJ}_{\mu}(v,z)}\frac{\delta g(v,y)}{\delta\pi^{\mu}_{IJ}(v,z)}
-\frac{\delta f(v,x)}{\delta\pi^{a}_{A}(v,z)}\frac{\delta g(v,y)}{\delta e^{A}_{a}(v,z)}
\nonumber \\&
-\frac{1}{2}\frac{\delta f(v,x)}{\delta\pi^{\mu}_{IJ}(v,z)}\frac{\delta g(v,y)}{\delta\omega^{IJ}_{\mu}(v,z)}]\od^{3}z,
\end{alignat}
$x$, $y$ and $z$ stand for 3-dimensional null hypersurface coordinates.
The Poisson brackets of canonical pairs are
\begin{alignat}{1}
\{e^{A}_{a}(v,x),\pi^{b}_{B}(v,y)\}=\delta^{A}_{B}\delta^{b}_{a}\delta^{3}(x-y),\quad
\{\omega^{IJ}_{\mu}(v,x),\pi^{\nu}_{KL}(v,y)\}=(\delta^{I}_{K}\delta^{J}_{L}-\delta^{I}_{L}\delta^{J}_{K})\delta^{\nu}_{\mu}\delta^{3}(x-y).
\end{alignat}

\section{Hamiltonian Analysis}

\setcounter{equation}{0}

\subsection{Total Hamiltonian}

By definition, the canonical momentum $P$ conjugate to a configuration variable $Q$ is
\begin{alignat}{1}
P:=\frac{\delta L}{\delta\dot{Q}},
\end{alignat}
and when the Lagrangian contains, at most, the linear term of $\dot{Q}$, the definition of the conjugate momentum $P$ gives a primary constraint.
Since the Palatini Lagrangian \eqref{2.22} is of the first order, one can get
\omits{Therefore, as to the Lagrangian \eqref{2.22}, one can get
\begin{alignat}{1}
\pi^a_{A}=0, \quad \pi^{0}_{-+}=&0, \quad \pi^1_{-+}=2\epsilon_{AB}\epsilon^{ab}e^A_a e^B_b,\quad \pi^a_{-+}=0,\quad
\pi^0_{-A}=0, \quad \pi^1_{-A}=0, \quad \pi^a_{-A}=4\epsilon_{AB}\epsilon^{ab}e^B_b, \nonumber \\
\pi^0_{+A}=&\pi^1_{+A}=\pi^a_{+A}=\pi^0_{23}=\pi^1_{23}=\pi^a_{23}=0,\label{3.2}
\end{alignat}
which should be treated as}28 primary constraints
\begin{alignat}{1}
\pi^a_{A}=0, \quad \pi^{0}_{-+}=&0, \quad \pi^1_{-+}-2\epsilon_{AB}\epsilon^{ab}e^A_a e^B_b=0,\quad \pi^a_{-+}=0,\quad \pi^0_{-A}=0, \quad \pi^1_{-A}=0, \nonumber \\
\pi^a_{-A}-4\epsilon_{AB}\epsilon^{ab}e^B_b=&0, \quad \pi^0_{+A}=\pi^1_{+A}=\pi^a_{+A}=\pi^0_{23}=\pi^1_{23}=\pi^a_{23}=0. \label{3.3}
\end{alignat}
Together with \eqref{2.9}, \eqref{2.10}, and \eqref{2.11}, there are 32 primary constraints in all.

By Legendre transformation, the canonical Hamiltonian is
\begin{alignat}{1}
H_{c}=&\int_{V}(\pi^{a}_{A}\dot{e}^{A}_{a}+\frac{1}{2}\pi^{\mu}_{IJ}\dot{\omega}^{IJ}_{\mu})\od^{3}x-\int_{V}\mathcal{L}\od^{3}x
\nonumber \\
=&\int_{V}\epsilon_{AB}\epsilon^{ab}[4(\omega^{-A}_{0,a}+\omega^{-+}_{0}\omega^{-A}_{a}-\omega^{-+}_{a}\omega^{-A}_{0}
-\omega^{-D}_{0}\omega^{CA}_{a}\delta_{DC}+\omega^{-D}_{a}\omega^{CA}_{0}\delta_{DC})e^B_b
\nonumber \\ &
+2(\omega^{-+}_{0,1}-\omega^{-C}_{1}\omega^{+D}_{0}\delta_{CD}+\omega^{-C}_{0}\omega^{+D}_{1}\delta_{CD})e^A_a e^B_b
-4F^{-+}_{1a}e^{A}_0e^{B}_b+4F^{-A}_{1a}l_0 e^{B}_b
\nonumber \\ &
-2F^{-A}_{ab}e^{B}_0-4n_{0}F^{+A}_{1a}e^{B}_b-n_{0}F^{AB}_{ab}]\od^3 x,
\end{alignat}
so the total Hamiltonian with primary constraints is
\begin{alignat}{1}
H_{T}=&\int_{V}[4\epsilon_{AB}\epsilon^{ab}(\omega^{-A}_{0,a}+\omega^{-+}_{0}\omega^{-A}_{a}-\omega^{-+}_{a}\omega^{-A}_{0}
-\omega^{-D}_{0}\omega^{CA}_{a}\delta_{DC}+\omega^{-D}_{a}\omega^{CA}_{0}\delta_{DC})e^B_b
\nonumber \\ &
+2\epsilon_{AB}\epsilon^{ab}(\omega^{-+}_{0,1}-\omega^{-C}_{1}\omega^{+D}_{0}\delta_{CD}
+\omega^{-C}_{0}\omega^{+D}_{1}\delta_{CD})e^A_a e^B_b
-n_{0}\epsilon_{AB}\epsilon^{ab}(4F^{+A}_{1a}e^{B}_b+F^{AB}_{ab})
\nonumber \\ &
+4\epsilon_{AB}\epsilon^{ab}F^{-A}_{1a}e^{B}_b l_0+e^{A}_0\epsilon_{AB}\epsilon^{ab}(2F^{-B}_{ab}-4F^{-+}_{1a}e^{B}_b)
+\lambda^{A}_{a}\pi^{a}_{A}+\lambda^{-+}_0\pi^0_{-+}
\nonumber \\ &
+\lambda^{-+}_1(\pi^1_{-+}-2\epsilon_{AB}\epsilon^{ab}e^A_a e^B_b)+\lambda^{-+}_a \pi^a_{-+}+\lambda^{-A}_0 \pi^0_{-A}
+\lambda^{-A}_1\pi^1_{-A}
+\lambda^{-A}_a(\pi^a_{-A}-4\epsilon_{AB}\epsilon^{ab}e^B_b)
\nonumber \\&
+\lambda^{+A}_0\pi^0_{+A}+\lambda^{+A}_1 \pi^1_{+A}+\lambda^{+A}_a \pi^a_{+A}
+\lambda^{23}_0\pi^0_{23}+\lambda^{23}_1 \pi^1_{23}+\lambda^{23}_a\pi^a_{23}]\od^3x.
\end{alignat}

\subsection{Consistency Analysis of Primary Constraints}

The primary constraints should always hold in the whole evolution. It means that their Poisson brackets with the total Hamiltonian should be zero
on the constraint surface in phase space. The following is the analysis of the consistency conditions for the primary constraints in details.
First, the consistency conditions for $\pi^a_A=0$ are
\begin{alignat}{1}
\{H_{T},\pi^{a}_{A}\}=&4\epsilon_{AB}\epsilon^{ab}(\omega^{-B}_{0,b}+\omega^{-+}_{0}\omega^{-B}_{b}-\omega^{-+}_{b}\omega^{-B}_{0}
-\omega^{-D}_{0}\omega^{CB}_{b}\delta_{DC}+\omega^{-D}_{b}\omega^{CB}_{0}\delta_{DC})
\nonumber \\ &
+4\epsilon_{AB}\epsilon^{ab} e^B_b(\omega^{-+}_{0,1}+\omega^{-C}_{1}\omega^{D+}_{0}\delta_{CD}
-\omega^{-C}_{0}\omega^{D+}_{1}\delta_{CD})-4\epsilon_{AB}\epsilon^{ab}e^B_b\lambda^{-+}_1
\nonumber \\ &
+4\epsilon_{AB}\epsilon^{ab}(l_0F^{-B}_{1b}-e^{B}_0F^{-+}_{1b}-n_{0}F^{+B}_{1b})
-4\epsilon_{AB}\epsilon^{ab}\lambda^{-B}_{b}\approx0. \label{19}
\end{alignat}
They will be always valid if
\begin{alignat}{1}
\lambda^{-+}_{1}\approx&\omega^{-+}_{0,1}-\omega^{-A}_{1}\omega^{+B}_{0}\delta_{AB}+\omega^{-A}_{0}\omega^{+B}_{1}\delta_{AB}+X^{-+}_{1},\\
\lambda^{-A}_{a}\approx&\omega^{-A}_{0,a}+\omega^{-+}_{0}\omega^{-A}_{a}-\omega^{-+}_{a}\omega^{-A}_{0}
-\omega^{-B}_{0}\omega^{CA}_{a}\delta_{BC}+\omega^{-B}_{a}\omega^{CA}_{0}\delta_{BC}
+e^{+}_0F^{-A}_{1a}-e^{-}_{0}F^{+A}_{1a}-e^{A}_0F^{-+}_{1a}-e^{A}_{a}X^{-+}_{1},\label{2.39}
\end{alignat}
where $X^{-+}_{1}$ is a function of canonical variables to be determined.

Next, the consistency conditions of the constraints with $\pi^\mu_{IJ}$ are
\begin{alignat}{1}
\{H_{T},\pi^{0}_{-+}\}=&-4\epsilon_{AB}\epsilon^{ab}e^{B}_{b}(e^{A}_{a,1}
+\omega^{AC}_{1}e^{D}_{a}\delta_{CD}-\omega^{-A}_{a})\approx0,\label{4.143}\\
\{H_{T},\pi^{1}_{-+}-2\epsilon_{AB}\epsilon^{ab}e^A_a e^B_b\}
=&-4\epsilon_{AB}\epsilon^{ab}e^{A}_{a}(e^{B}_{0,b}+\omega^{+B}_{b}n_{0}+\omega^{-B}_{b}l_{0}
+\omega^{BC}_{b}e^{D}_{0}\delta_{CD}-\omega^{BC}_{0}e^{D}_{b}\delta_{CD})
\nonumber \\&
+4\epsilon_{AB}\epsilon^{ab}e^{B}_{0}(e^{A}_{b,a}+\omega^{AC}_{a}e^{D}_{b}\delta_{CD})
+4\epsilon_{AB}\epsilon^{ab}\lambda^{A}_{a}e^{B}_{b}\approx0,\\
\{H_{T},\pi^{a}_{-+}\}=&4\epsilon_{AB}\epsilon^{ab}[e^{B}_{b}(e^{A}_{0,1}-\omega^{-A}_{0}+\omega^{+A}_{1}n_{0}
+\omega^{-A}_{1}l_0+\omega^{AC}_{1}e^{D}_{0}\delta_{CD})
\nonumber \\&
-e^{B}_0(e^{A}_{b,1}-\omega^{-A}_{b}+\omega^{AC}_{1}e^{D}_{b}\delta_{CD})]\approx0,\\
\{H_{T},\pi^{0}_{-A}\}=&4\epsilon_{AB}\epsilon^{ab}e^{B}_{b}(\omega^{+C}_{1}e^{D}_{a}\delta_{CD}-\omega^{-+}_{a})
+4\epsilon_{AB}\epsilon^{ab}(e^{B}_{a,b}+\omega^{BC}_{b}e^{D}_{a}\delta_{CD})
\approx0,\\\label{4.146}
\{H_{T},\pi^{1}_{-A}\}=&-4\epsilon_{AB}\epsilon^{ab}e^{B}_{0}(\omega^{+C}_{a}e^{D}_b\delta_{CD})
-4\epsilon_{AB}\epsilon^{ab}l_{0}(e^{B}_{a,b}+\omega^{BC}_{b}e^{D}_{a}\delta_{CD})
\nonumber \\ &
+4\epsilon_{AB}\epsilon^{ab} e^B_b(l_{0,a}+\omega^{-+}_{a}l_{0}-\omega^{+C}_{0}e^D_a\delta_{CD}+\omega^{+C}_{a}e^{D}_0\delta_{CD})\approx0,\\
\{H_{T},\pi^{a}_{-A}-4\epsilon_{AB}\epsilon^{ab}e^B_b\}
=&4\epsilon_{AB}\epsilon^{ab}e^B_b\omega^{-+}_{0}+4\epsilon_{AB}\epsilon^{ab}\omega^{BC}_{0}e^D_b\delta_{CD}
+4\epsilon_{AB}\epsilon^{ab}e^{B}_0\omega^{+C}_{1}e^{D}_b\delta_{CD}
\nonumber \\ &
-4\epsilon_{AB}\epsilon^{ab}e^{B}_b\omega^{+C}_{1}e^{D}_0\delta_{CD}
-4\epsilon_{AB}\epsilon^{ab}(l_0 e^{B}_b)_{,1}+4\epsilon_{CA}\epsilon^{ab}l_0 e^{C}_b\omega^{-+}_{1}
\nonumber \\ &
-4\epsilon_{AB}\epsilon^{ab}l_{0}\omega^{BC}_{1}e^{D}_b\delta_{CD}
-4\epsilon_{AB}\epsilon^{ab}(e^{B}_{0})_{,b}-4\epsilon_{AB}\epsilon^{ab}\omega^{BC}_{b}e^{D}_0\delta_{CD}
\nonumber \\ &
-4\epsilon_{AB}\epsilon^{ab}e^{B}_0\omega^{-+}_{b}-4\epsilon_{AB}\epsilon^{ab}\omega^{+B}_{b}n_{0}
+4\epsilon_{AB}\epsilon^{ab}\lambda^{B}_{b}\approx0,
\end{alignat}
\begin{alignat}{1}
\{H_{T},\pi^{0}_{+A}\}=&-2\epsilon_{BC}\epsilon^{ab}e^{B}_{a}e^{C}_{b}\omega^{-A}_{1}=-4e\omega^{-A}_{1}\approx0,\label{4.149}\\
\{H_{T},\pi^{1}_{+A}\}=&4\epsilon_{AB}\epsilon^{ab}n_{0}(e^{B}_{a,b}+\omega^{BC}_{b}e^{D}_{a}\delta_{CD})
+4\epsilon_{AB}\epsilon^{ab}e^{B}_{0}\omega^{-C}_{a}e^D_b\delta_{CD}
\nonumber \\&
-4\epsilon_{AB}\epsilon^{ab}e^B_a(-n_{0,b}+\omega^{-C}_{0}e^{D}_{b}\delta_{CD}+\omega^{-+}_{b}n_{0}-\omega^{-C}_{b}e^{D}_{0}\delta_{CD})\approx0,\\
\{H_{T},\pi^{a}_{+A}\}=&-4\epsilon_{AB}\epsilon^{ab}(-n_{0,1}e^{B}_{b}-n_{0}e^{B}_{b,1}
-\omega^{BC}_{1}n_{0}e^{D}_{b}\delta_{CD}+\omega^{-B}_{b}n_{0})-4\epsilon_{AB}\epsilon^{ab}e^{B}_b\omega^{-+}_{1}n_{0}
\nonumber \\&
+4\epsilon_{BC}\epsilon^{ab}e^{B}_0e^{C}_b\omega^{-D}_{1}\delta_{DA}
\nonumber \\
=&-4\epsilon_{AB}\epsilon^{ab}e^{B}_b\omega^{-+}_{1}n_{0}+4\epsilon_{AB}\epsilon^{ab}n_{0,1}e^{B}_{b}
+4\epsilon_{BC}\epsilon^{ab}e^{B}_0e^{C}_b\omega^{-D}_{1}\delta_{DA}
\nonumber \\&
+4\epsilon_{AB}\epsilon^{ab}n_{0}(e^{B}_{b,1}+\omega^{BC}_{1}e^{D}_{b}\delta_{CD}-\omega^{-B}_{b})\approx0,\\
\{H_{T},\pi^{0}_{23}\}=&-4\epsilon^{ab}\omega^{-C}_{a}e^D_b\delta_{CD}\approx0,\label{4.152}\\
\{H_{T},\pi^{1}_{23}\}=&4\epsilon^{ab}l_0\omega^{-C}_{a}e^{D}_b\delta_{CD}
-4\epsilon^{ab}n_{0}\omega^{+C}_{a}e^{D}_b\delta_{CD}\approx0,\\
\{H_{T},\pi^{a}_{23}\}=&4\epsilon^{ab}(n_{0,b}+\omega^{-C}_{0} e^D_b\delta_{CD}-\omega^{-C}_{b}e^{D}_0\delta_{CD}+\omega^{-+}_{b}n_{0})
-4\epsilon^{ab}n_{0}(\omega^{-+}_{b}-\omega^{+C}_{1}e^{D}_{b}\delta_{CD})
\nonumber \\
\approx&0.
\end{alignat}
The above 24 conditions are equal to 24 torsion-free conditions
\begin{alignat}{1}
&n_{0,1}-\omega^{-+}_{1}n_{0}\approx0,\label{2.24}\\
&n_{0,a}-\omega^{-+}_{a}n_{0}-\omega^{-A}_{0}e^{B}_{a}\delta_{AB}+\omega^{-A}_{a}e^{B}_{0}\delta_{AB}\approx0,\label{2.31}\\
&l_{0,1}-\omega^{-+}_{0}+\omega^{+A}_{1}e^{B}_0\delta_{AB}+\omega^{-+}_{1}l_{0}\approx0,\label{2.25}\\
&l_{0,a}+\omega^{+A}_{a}e^{B}_0\delta_{AB}-\omega^{+A}_{0}e^{B}_a\delta_{AB}+\omega^{-+}_{a}l_{0}\approx0,\label{2.26}\\
&e^{A}_{0,1}-\omega^{-A}_{0}+\omega^{+A}_{1}n_{0}+\omega^{AB}_{1}e^{C}_{0}\delta_{BC}\approx0,\label{2.27}\\
&e^{A}_{0,a}-\lambda^{A}_{a}+\omega^{-A}_{a}l_0+\omega^{+A}_{a}n_{0}+\omega^{AB}_{a}e^{C}_0\delta_{BC}
-\omega^{AB}_{0}e^C_a\delta_{BC}\approx0,\label{4.173}
\end{alignat}
and
\begin{alignat}{1}
&\omega^{-A}_{1}\approx0,\label{2.32}\\
&\epsilon^{ab}\omega^{-A}_{a}e^B_b\delta_{AB}\approx0,\label{5.158}\\
&\omega^{-+}_{a}-\omega^{+A}_{1}e^{B}_{a}\delta_{AB}\approx0,\label{2.30}\\
&\epsilon^{ab}\omega^{+A}_{a}e^{B}_b\delta_{AB}\approx0,\label{2.31-1}\\
&e^{A}_{a,1}-\omega^{-A}_{a}+\omega^{AB}_{1}e^{C}_{a}\delta_{BC}\approx0,\label{5.162}\\
&\epsilon^{ab}(e^{A}_{a,b}-\omega^{AB}_{a}e^{C}_{b}\delta_{BC})\approx0.\label{2.33}
\end{alignat}
\eqref{4.173} are 4 torsion-free conditions by using the equations of motion of $e^{A}_{a}$
\begin{alignat}{1}
\dot{e}^{A}_{a}=\{e^{A}_{a},H_{T}\}=\lambda^{A}_{a}\approx&e^{A}_{0,a}+\omega^{-A}_{a}l_0+\omega^{+A}_{a}n_{0}+\omega^{AB}_{a}e^{C}_0\delta_{BC}
-\omega^{AB}_{0}e^C_a\delta_{BC},
\end{alignat}
which result in
\begin{alignat}{1}
e^{A}_{0,a}-e^{A}_{a,0}+\omega^{-A}_{a}l_0+\omega^{+A}_{a}n_{0}+\omega^{AB}_{a}e^{C}_0\delta_{BC}
-\omega^{AB}_{0}e^C_a\delta_{BC}\approx0. \label{3.33}
\end{alignat}
The last 12 torsion-free conditions \eqref{2.32}-\eqref{2.33} contain no multipliers,
so they are 12 secondary constraints.

Finally, the consistency conditions for \eqref{2.9}, \eqref{2.10} and \eqref{2.11} are as follows:
\begin{alignat}{1}
\{\epsilon_{AB}\epsilon^{ab}F^{-A}_{1a}e^{B}_b,H_{T}\}
=&\epsilon_{AB}\epsilon^{ab}\{F^{-A}_{1a},H_{T}\}e^{B}_b+\epsilon_{AB}\epsilon^{ab}F^{-A}_{1a}\{e^{B}_b,H_{T}\}
\nonumber \\
=&\epsilon_{AB}\epsilon^{ab}(\lambda^{-A}_{a,1}-\lambda^{-+}_{1}\omega^{-A}_{a}-\omega^{-+}_{1}\lambda^{-A}_{a}
-\lambda^{-C}_{a}\omega^{DA}_{1}\delta_{CD})e^{B}_{b}+\epsilon_{AB}\epsilon^{ab}F^{-A}_{1a}\lambda^{B}_b\approx0,\label{2.62}
\end{alignat}
which will be a trivial identity after the determination of $\lambda^{-+}_{1}$ and $\lambda^{-A}_{a}$, see appendix A.
\begin{alignat}{1}
\{\epsilon_{AB}\epsilon^{ab}F^{+A}_{1a}e^{B}_{b}+F^{23}_{23},H_{T}\}
=\epsilon_{AB}\epsilon^{ab}\{F^{+A}_{1a},H_{T}\}e^{B}_{b}+\epsilon_{AB}\epsilon^{ab}F^{+A}_{1a}\lambda^{B}_{b}+\{F^{23}_{23},H_{T}\}\approx0,
\label{2.37}
\end{alignat}
where
\begin{alignat}{1}
\{F^{+A}_{1a},H_{T}\}=&\lambda^{+A}_{a,1}-\lambda^{+A}_{1,a}+\lambda^{-+}_{1}\omega^{+A}_{a}+\omega^{-+}_{1}\lambda^{+A}_{a}
+\lambda^{+B}_{1}\omega^{CA}_{a}\delta_{BC}+\omega^{+B}_{1}\lambda^{CA}_{a}\delta_{BC}
\nonumber \\&
-\lambda^{-+}_{a}\omega^{+A}_{1}-\omega^{-+}_{a}\lambda^{+A}_{1}-\lambda^{+B}_{a}\omega^{CA}_{1}\delta_{BC}-\omega^{+B}_{a}\lambda^{CA}_{1}\delta_{BC},\\
\{F^{23}_{23},H_{T}\}=&\lambda^{23}_{3,2}-\lambda^{23}_{2,3}+\epsilon^{ab}\lambda^{+2}_{a}\omega^{-3}_{b}
+\epsilon^{ab}\omega^{+2}_{a}\lambda^{-3}_{b}+\epsilon^{ab}\lambda^{-2}_{a}\omega^{+3}_{b}
+\epsilon^{ab}\omega^{-2}_{a}\lambda^{+3}_{b}.
\end{alignat}
\begin{alignat}{1}
\{F^{-A}_{23}+\epsilon^{ab}e^{A}_aF^{-+}_{1b},H_{T}\}
=\{F^{-A}_{23},H_{T}\}+\epsilon^{ab}\lambda^{A}_aF^{-+}_{1b}+\epsilon^{ab}e^{A}_a\{F^{-+}_{1b},H_{T}\}
\approx0,\label{2.40}
\end{alignat}
where
\begin{alignat}{1}
\{F^{-A}_{23},H_{T}\}=&-\epsilon^{ab}\lambda^{-A}_{a,b}-\epsilon^{ab}\lambda^{-+}_{a}\omega^{-A}_{b}-\epsilon^{ab}\omega^{-+}_{a}\lambda^{-A}_{b}
+\epsilon^{ab}\lambda^{-B}_{a}\omega^{CA}_{b}\delta_{BC}+\epsilon^{ab}\omega^{-B}_{a}\lambda^{CA}_{b}\delta_{BC},\\
\{F^{-+}_{1b},H_{T}\}=&\lambda^{-+}_{b,1}-\lambda^{-+}_{1,b}+\lambda^{-B}_{b}\omega^{+C}_{1}\delta_{BC}+\omega^{-B}_{b}\lambda^{+C}_{1}\delta_{BC}.
\end{alignat}
\eqref{2.37} and \eqref{2.40} set 3 relations among the multipliers.

\subsection{Consistency Analysis of Secondary Constraints}

The secondary constraints should also preserve in the evolution, which requires
\begin{alignat}{1}
&\{\omega^{-A}_{1},H_{T}\}=\lambda^{-A}_{1}\approx0,\\
&\{\epsilon^{ab}\omega^{-A}_{a}e^B_b\delta_{AB},H_{T}\}=\epsilon^{ab}\lambda^{-A}_{a}e^B_b\delta_{AB}
+\epsilon^{ab}\omega^{-A}_{a}\lambda^B_b\delta_{AB}\approx0,\label{3.46}\\
&\{\omega^{-+}_{a}-\omega^{+A}_{1}e^{B}_{a}\delta_{AB},H_{T}\}
=\lambda^{-+}_{a}-\lambda^{+A}_{1}e^{B}_{a}\delta_{AB}-\omega^{+A}_{1}\lambda^{B}_{a}\delta_{AB}\approx0,\label{2.48}\\
&\{\epsilon^{ab}\omega^{+A}_{a}e^{B}_b\delta_{AB},H_{T}\}
=\epsilon^{ab}\lambda^{+A}_{a}e^{B}_b\delta_{AB}+\epsilon^{ab}\omega^{+A}_{a}\lambda^{B}_b\delta_{AB}\approx0,\\
&\{e^{A}_{a,1}+\omega^{AB}_{1}e^{C}_{a}\delta_{BC}-\omega^{-A}_{a},H_{T}\}
=\lambda^{A}_{a,1}+\lambda^{AB}_{1}e^{C}_{a}\delta_{BC}+\omega^{AB}_{1}\lambda^{C}_{a}\delta_{BC}-\lambda^{-A}_{a}
\nonumber \\&
\approx(\lambda^{AB}_{1}-\omega^{AB}_{0,1}+\omega^{-A}_{0}\omega^{+B}_{1}-\omega^{+A}_{1}\omega^{-B}_{0})e^{C}_{a}\delta_{CB}
+2n_{0}F^{+A}_{1a}+F^{AB}_{1a}e^{C}_{0}\delta_{BC}+e^{A}_0F^{-+}_{1a}+e^{A}_{a}X^{-+}_{1}\approx0,\label{3.49}\\
&\{\epsilon^{ab}(e^{A}_{a,b}-\omega^{AB}_{a}e^{C}_{b}\delta_{BC}),H_{T}\}
=\epsilon^{ab}\lambda^{A}_{a,b}-\epsilon^{ab}\lambda^{AB}_{a}e^{C}_{b}\delta_{BC}-\epsilon^{ab}\omega^{AB}_{a}\lambda^{C}_{b}\delta_{BC}\approx0.\label{3.50}
\end{alignat}

Combined with \eqref{2.39}, \eqref{4.173}, and \eqref{3.46}, one can get
\begin{alignat}{1}
&\epsilon^{ab}(\omega^{-A}_{0,a}-\omega^{-+}_{a}\omega^{-A}_{0}-\omega^{-B}_{0}\omega^{CA}_{a}\delta_{BC}
+\omega^{-B}_{a}\omega^{CA}_{0}\delta_{BC})e^D_b\delta_{AD}
+\epsilon^{ab}(l_0F^{-A}_{1a}-e^{A}_0F^{-+}_{1a}-n_{0}F^{+A}_{1a})e^B_b\delta_{AB}
\nonumber \\ &
-\epsilon^{ab}(e^{A}_{0,a}+\omega^{-A}_{a}l_0+\omega^{+A}_{a}n_{0}+\omega^{AB}_{a}e^{C}_0\delta_{BC}-\omega^{AB}_{0}e^C_a\delta_{BC})
\omega^{-D}_{b}\delta_{AD}\approx0, \label{2.47}
\end{alignat}
which will be automatically satisfied after the determination of $X^{-+}_{1}$, see appendix B.

\eqref{3.49} leads to 4 expressions of $\lambda^{23}_{1}$:
\begin{alignat}{1}
\lambda^{23}_{1(1)}\approx&\omega^{23}_{0,1}-\omega^{-2}_{0}\omega^{+3}_{1}+\omega^{-3}_{0}\omega^{+2}_{1}-2n_{0}F^{+2}_{12}(e^{3}_{2})^{-1}
-F^{23}_{12}e^{3}_{0}(e^{3}_{2})^{-1}-e^{2}_0F^{-+}_{12}(e^{3}_{2})^{-1}-e^{2}_{2}(e^{3}_{2})^{-1}X^{-+}_{1},\label{3.48}\\
\lambda^{23}_{1(2)}\approx&\omega^{23}_{0,1}-\omega^{-2}_{0}\omega^{+3}_{1}+\omega^{+2}_{1}\omega^{-3}_{0}-2n_{0}F^{+2}_{13}(e^{3}_{3})^{-1}
-F^{23}_{13}e^{3}_{0}(e^{3}_{3})^{-1}-e^{2}_0F^{-+}_{13}(e^{3}_{3})^{-1}-e^{2}_{3}(e^{3}_{3})^{-1}X^{-+}_{1},\\
\lambda^{23}_{1(3)}\approx&\omega^{23}_{0,1}-\omega^{-2}_{0}\omega^{+3}_{1}+\omega^{+2}_{1}\omega^{-3}_{0}+2n_{0}F^{+3}_{12}(e^{2}_{2})^{-1}
-F^{23}_{12}e^{2}_{0}(e^{2}_{2})^{-1}+e^{3}_0F^{-+}_{12}(e^{2}_{2})^{-1}+e^{3}_{2}(e^{2}_{2})^{-1}X^{-+}_{1},\\
\lambda^{23}_{1(4)}\approx&\omega^{23}_{0,1}-\omega^{-2}_{0}\omega^{+3}_{1}+\omega^{+2}_{1}\omega^{-3}_{0}+2n_{0}F^{+3}_{13}(e^{2}_{3})^{-1}
-F^{23}_{13}e^{2}_{0}(e^{2}_{3})^{-1}+e^{3}_0F^{-+}_{13}(e^{2}_{3})^{-1}+e^{3}_{3}(e^{2}_{3})^{-1}X^{-+}_{1}.
\end{alignat}
They should be equal to each other. From them, one can get 2 new secondary constraints:
\begin{alignat}{1}
\epsilon^{ab}F^{+2}_{1a}e^{3}_{b}+\epsilon^{ab}F^{+3}_{1a}e^{2}_{b}\approx&0,\label{3.56}\\
\epsilon^{ab}F^{+2}_{1a}e^{2}_{b}-\epsilon^{ab}F^{+3}_{1a}e^{3}_{b}\approx&0,\label{3.57}
\end{alignat}
and determine $X^{-+}_{1}$ as
\begin{alignat}{1}
X^{-+}_{1}\approx n_{0}F^{23}_{23}e^{-1}-\epsilon_{AB}\epsilon^{ab}e^{A}_{0}F^{-+}_{1a}e^{B}_{b}e^{-1}.
\end{alignat}
Therefore,
\begin{alignat}{1}
\lambda^{-+}_{1}\approx&\omega^{-+}_{0,1}+\omega^{-A}_{0}\omega^{+B}_{1}\delta_{AB}+n_{0}F^{23}_{23}e^{-1}
-\epsilon_{AB}\epsilon^{ab}e^{A}_{0}F^{-+}_{1a}e^{B}_{b}e^{-1}=:\Lambda^{-+}_{1},\label{3.55}\\
\lambda^{-A}_{a}\approx&\omega^{-A}_{0,a}+\omega^{-+}_{0}\omega^{-A}_{a}-\omega^{-+}_{a}\omega^{-A}_{0}
-\omega^{-B}_{0}\omega^{CA}_{a}\delta_{BC}+\omega^{-B}_{a}\omega^{CA}_{0}\delta_{BC}+e^{+}_0F^{-A}_{1a}
\nonumber \\ &
-e^{-}_{0}F^{+A}_{1a}-e^{A}_0F^{-+}_{1a}-n_{0}e^{A}_{a}F^{23}_{23}e^{-1}
+e^{A}_{a}\epsilon_{BC}\epsilon^{bc}e^{B}_{0}F^{-+}_{1b}e^{C}_{c}e^{-1}
=:\Lambda^{-A}_{a}.\label{3.60}
\end{alignat}
From \eqref{3.50}, the multipliers $\lambda^{23}_a$ can be determined:
\begin{alignat}{1}
\lambda^{23}_{a}\approx&\omega^{23}_{0,a}-\omega^{-2}_{0}\omega^{+3}_{a}+\omega^{-3}_{0}\omega^{+2}_{a}
-\omega^{+2}_{0}\omega^{-3}_{a}+\omega^{+3}_{0}\omega^{-2}_{a}-n_{0}e^{A}_{a}F^{+B}_{23}\delta_{AB}e^{-1}
-l_0e^{A}_{a}F^{-B}_{23}\delta_{AB}e^{-1}
\nonumber \\ &
-\epsilon_{AB}e^{A}_{a}e^{B}_0F^{23}_{23}e^{-1}
=:\Lambda^{23}_{a}.\label{3.61}
\end{alignat}

\subsection{Consistency Analysis of Further Secondary Constraints}

The consistency conditions of the further secondary constraints \eqref{3.56} and \eqref{3.57} are
\begin{alignat}{1}
\{\epsilon^{ab}F^{+2}_{1a}e^{3}_{b}+\epsilon^{ab}F^{+3}_{1a}e^{2}_{b},H_{T}\}
=&\epsilon^{ab}\{F^{+2}_{1a},H_{T}\}e^{3}_{b}+\epsilon^{ab}F^{+2}_{1a}\lambda^{3}_{b}
+\epsilon^{ab}\{F^{+3}_{1a},H_{T}\}e^{2}_{b}+\epsilon^{ab}F^{+3}_{1a}\lambda^{2}_{b}\approx0,\label{2.60}\\
\{\epsilon^{ab}F^{+2}_{1a}e^{2}_{b}-\epsilon^{ab}F^{+3}_{1a}e^{3}_{b},H_{T}\}
=&\epsilon^{ab}\{F^{+2}_{1a},H_{T}\}e^{2}_{b}+\epsilon^{ab}F^{+2}_{1a}\lambda^{2}_{b}
-\epsilon^{ab}\{F^{+3}_{1a},H_{T}\}e^{3}_{b}-\epsilon^{ab}F^{+3}_{1a}\lambda^{3}_{b}\approx0, \label{2.61}
\end{alignat}
where
\begin{alignat}{1}
\{F^{+2}_{1a},H_{T}\}=&\lambda^{+2}_{a,1}-\lambda^{+2}_{1,a}+\lambda^{-+}_{1}\omega^{+2}_{a}+\omega^{-+}_{1}\lambda^{+2}_{a}
+\lambda^{+3}_{1}\omega^{32}_{a}+\omega^{+3}_{1}\lambda^{32}_{a}-\lambda^{-+}_{a}\omega^{+2}_{1}
-\omega^{-+}_{a}\lambda^{+2}_{1}-\lambda^{+3}_{a}\omega^{32}_{1}-\omega^{+3}_{a}\lambda^{32}_{1}
\nonumber \\
\approx&\lambda^{+2}_{a,1}-\lambda^{+2}_{1,a}+\Lambda^{-+}_{1}\omega^{+2}_{a}+\omega^{-+}_{1}\lambda^{+2}_{a}
+\lambda^{+3}_{1}\omega^{32}_{a}+\omega^{+3}_{1}\Lambda^{32}_{a}-\lambda^{-+}_{a}\omega^{+2}_{1}
-\omega^{-+}_{a}\lambda^{+2}_{1}-\lambda^{+3}_{a}\omega^{32}_{1}-\omega^{+3}_{a}\Lambda^{32}_{1},\\
\{F^{+3}_{1a},H_{T}\}=&\lambda^{+3}_{a,1}-\lambda^{+3}_{1,a}+\lambda^{-+}_{1}\omega^{+3}_{a}+\omega^{-+}_{1}\lambda^{+3}_{a}
+\lambda^{+2}_{1}\omega^{23}_{a}+\omega^{+2}_{1}\lambda^{23}_{a}-\lambda^{-+}_{a}\omega^{+3}_{1}
-\omega^{-+}_{a}\lambda^{+3}_{1}-\lambda^{+2}_{a}\omega^{23}_{1}-\omega^{+2}_{a}\lambda^{23}_{1}
\nonumber \\
\approx&\lambda^{+3}_{a,1}-\lambda^{+3}_{1,a}+\Lambda^{-+}_{1}\omega^{+3}_{a}+\omega^{-+}_{1}\lambda^{+3}_{a}
+\lambda^{+2}_{1}\omega^{23}_{a}+\omega^{+2}_{1}\Lambda^{23}_{a}-\lambda^{-+}_{a}\omega^{+3}_{1}
-\omega^{-+}_{a}\lambda^{+3}_{1}-\lambda^{+2}_{a}\omega^{23}_{1}-\omega^{+2}_{a}\Lambda^{23}_{1}.
\end{alignat}
They are relations among the multipliers.

\subsection{Integrability}

Eqs.\eqref{2.24}-\eqref{4.173} define the first derivatives of $n_0$, $l_0$ and $e^A_0$ with respect to their spatial coordinates $x^1$ and $x^a$.
As a self-consistent system, these multipliers ($n_0$, $l_0$ and $e^A_0$) should satisfy the integrability conditions.
Therefore, we should check whether the integrability conditions will result in new constraints.
The direct calculations show that all the integrability conditions result in the Ricci identities.
The detailed calculation will be left in Appendix C, D and E, respectively.

\subsection{Equations of Motion}

The equations of motion of the configuration variables are
\begin{alignat}{1}
\dot{e}^{A}_{a}=&\{e^{A}_{a},H_{T}\}=\lambda^{A}_{a}\approx e^{A}_{0,a}+\omega^{-A}_{a}l_0+\omega^{+A}_{a}n_{0}
+\omega^{AB}_{a}e^{C}_0\delta_{BC}-\omega^{AB}_{0}e^C_a\delta_{BC},\\
\dot{\omega}^{-+}_{0}=&\{\omega^{-+}_{0},H_{T}\}=\lambda^{-+}_{0},\quad
\dot{\omega}^{-+}_{1}=\{\omega^{-+}_{1},H_{T}\}=\lambda^{-+}_{1},\quad
\dot{\omega}^{-+}_{a}=\{\omega^{-+}_{a},H_{T}\}=\lambda^{-+}_{a},\label{3.69}\\
\dot{\omega}^{-A}_{0}=&\{\omega^{-A}_{0},H_{T}\}=\lambda^{-A}_{0},\quad
\dot{\omega}^{-A}_{1}=\{\omega^{-A}_{1},H_{T}\}=\lambda^{-A}_{1}\approx 0,\quad
\dot{\omega}^{-A}_{a}=\{\omega^{-A}_{a},H_{T}\}=\lambda^{-A}_{a},\\
\dot{\omega}^{+A}_{0}=&\{\omega^{+A}_{0},H_{T}\}=\lambda^{+A}_{0},\quad
\dot{\omega}^{+A}_{1}=\{\omega^{+A}_{1},H_{T}\}=\lambda^{+A}_{1},\quad
\dot{\omega}^{+A}_{a}=\{\omega^{+A}_{a},H_{T}\}=\lambda^{+A}_{a},\label{3.75}\\
\dot{\omega}^{23}_{0}=&\{\omega^{23}_{0},H_{T}\}=\lambda^{23}_{0},\quad
\dot{\omega}^{23}_{1}=\{\omega^{23}_{1},H_{T}\}=\lambda^{23}_{1},\quad
\dot{\omega}^{23}_{a}=\{\omega^{23}_{a},H_{T}\}=\lambda^{23}_{a}.
\end{alignat}
The equations of motion of the non-vanishing conjugate momenta are
\begin{alignat}{1}
\dot{\pi}^{1}_{-+}=\{\pi^{1}_{-+},H_{T}\}\approx4\epsilon_{AB}\epsilon^{ab}\lambda^A_a e^B_b,\quad
\dot{\pi}^{a}_{-A}=\{\pi^{a}_{-A},H_{T}\}\approx4\epsilon_{AB}\epsilon^{ab}\lambda^B_b.
\end{alignat}

\section{Classifications of Constraints}

\setcounter{equation}{0}

\subsection{First and Second Class Constraints}

One can see that there are 6 first class constraints
\begin{alignat}{1}
\pi^{0}_{-+}=0,\quad \pi^{0}_{-A}=0,\quad \pi^{0}_{+A}=0,\quad \pi^{0}_{23}=0,
\end{alignat}
because their corresponding configuration variables $\omega^{IJ}_{0}$ do not exist in the constraints.
The left 40 constraints are of the second class. The Poisson brackets of the constraints can be found at appendix F.

\subsection{Degrees of Freedom}

There are 4+24=28 configuration variables and 28 conjugate momenta in this system, which span a 56-dimensional phase space.
There are 46 constraints, including 32 primary constraints and 14 secondary constraints.
Among the 46 constraints, there are 6 first class constraints, and 40 second class constraints,
which altogether reduce 52 degrees of freedom in the phase space. Therefore, there are 4 degrees of freedom left in the phase space,
which means there are 2 local physical degrees of freedom. They correspond to 2 independent polarization modes of the gravitational wave.

\subsection{Scalar and Vector Constraints}

In $\mathfrak{su}$(2)-connection dynamics \cite{Ashtekar}, the constraints are classified as the spatial scalar, spatial vector
and $\mathfrak{su}$(2) gauge constraints. In comparison, $\epsilon_{AB}\epsilon^{ab}F^{+A}_{1a}e^{B}_{b}+F^{23}_{23}\approx0$ and $\epsilon_{AB}\epsilon^{ab}F^{-A}_{1a}e^{B}_b\approx0$ are two scalar constraints and $F^{-A}_{23}+\epsilon^{ab}e^{A}_aF^{-+}_{1b}\approx0$
is a 2-dimensional vector constraint. The vector constraint reduces 2 degrees of freedom in the phase space,
because it is actually composed of 2 second class constraints.

\subsection{Gauss Constraints}

In the new approach, the Gauss constraints are not independent ones. They can be read out in the above analysis as the following way.

The SO(1,3) Gauss constraints can be written as \cite{BTT}
\begin{alignat}{1}
D_{j}\pi^{j}_{IJ}:=\partial_{j}\pi^{j}_{IJ}+\eta_{IK}\omega^{KL}_{j}\pi^{j}_{LJ}-\eta_{JK}\omega^{KL}_{j}\pi^{j}_{LI}\approx0.\label{3.65-1}
\end{alignat}
By using primary constraints \eqref{3.3} to replace coframe $e^{A}_{a}$ by no-zero conjugate momenta, one can see that the above constraints (\ref{3.65-1})
are actually the consistency conditions of the 6 primary constraints $\pi^{0}_{IJ}=0$.

The SO(1,1) gauge constraint comes from the consistency condition of $\pi^{0}_{-+}=0$:
\begin{alignat}{1}
\{H_{T},\pi^{0}_{-+}\}=&-4\epsilon_{AB}\epsilon^{ab}e^{B}_{b}(e^{A}_{a,1}+\omega^{AC}_{1}e^{D}_{a}\delta_{CD}-\omega^{-A}_{a})
\approx-\partial_{j}\pi^{j}_{-+}-\eta_{-K}\omega^{KL}_{j}\pi^{j}_{L+}+\eta_{+K}\omega^{KL}_{j}\pi^{j}_{L-}
\nonumber \\
=&-D_{j}\pi^{j}_{-+}\approx0.\label{5.48}
\end{alignat}

The T$^{-}$(2) gauge constraints come from the consistency conditions of $\pi^{0}_{-A}=0$:
\begin{alignat}{1}
\{H_{T},\pi^{0}_{-A}\}=&4\epsilon_{AB}\epsilon^{ab}(e^{B}_{a,b}+\omega^{BC}_{b}e^{D}_{a}\delta_{CD})
+4\epsilon_{AB}\epsilon^{ab}e^{B}_{b}(\omega^{+C}_{1}e^{D}_{a}\delta_{CD}-\omega^{-+}_{a})
\nonumber \\
\approx&-\partial_{j}\pi^{j}_{-A}-\eta_{-K}\omega^{KL}_{j}\pi^{j}_{LA}+\eta_{AK}\omega^{KL}_{j}\pi^{j}_{L-}
=-D_{j}\pi^{j}_{-A}\approx0.\label{3.91}
\end{alignat}

The T$^{+}$(2) gauge constraints come from the consistency conditions of $\pi^{0}_{+A}=0$:
\begin{alignat}{1}
\{H_{T},\pi^{0}_{+A}\}=-2\epsilon_{CB}\epsilon^{ab}e^{C}_{a}e^{B}_{b}\omega^{-A}_{1}=-4e\omega^{-A}_{1}
\approx-\partial_{j}\pi^{j}_{+A}-\eta_{+K}\omega^{KL}_{j}\pi^{j}_{LA}+\eta_{AK}\omega^{KL}_{j}\pi^{j}_{L+}
=-D_{j}\pi^{j}_{+A}\approx0.\label{5.50}
\end{alignat}

The SO(2) gauge constraint comes from the consistency condition of $\pi^{0}_{23}=0$:
\begin{alignat}{1}
\{H_{T},\pi^{0}_{23}\}=&-4\epsilon^{ab}\omega^{-C}_{a}e^D_b\delta_{CD}
\approx-\partial_{j}\pi^{j}_{23}-\eta_{2K}\omega^{KL}_{j}\pi^{j}_{L3}+\eta_{3K}\omega^{KL}_{j}\pi^{j}_{L2}
=-D_{j}\pi^{j}_{23}\approx0.\label{5.51}
\end{alignat}

\section{Summary}

A self-consistent Hamiltonian formalism for a 4-dimensional connection dynamics has been set up in a Bondi-like coordinate system $\{v,r,x^{a}\}$.
The advanced null coordinate $v$ is used as the time coordinate instead of $u$ in the Bondi-Sachs coordinates.
3 components of the metric are fixed in the Bondi-like metric, so there are only 7 nonzero components in the metric.
The 3 Bondi-like conditions can be translated into 3 conditions on the coframe, and can be treated as 3 primary constraints as well,
which will preserve in the evolution. The 3-dimensional hypersurfaces labelled by $v$ have degenerate metric,
so they are null hypersurfaces.

The internal symmetry SO(1,3) is decomposed to SO(1,1), SO(2), T$^{\pm}(2)$,
and the Lie algebra $\mathfrak{so}$(1,3) is spanned by $\{L_{-+}, L_{23}, L_{-A}, L_{+A}\}$.
The coframe consists of 2 null 1-forms and 2 spacelike 1-forms. A simple coframe has been chosen to make 
Hamiltonian analysis.  The $\mathfrak{so}$(1,3) connection has 24 components, which are treated as 24 independent configuration variables.
They together with 4 coframe coefficients $e^A_a$ and their conjugate momenta span a 56-dimensional phase space.
There are 32 primary constraints and 14 secondary constraints. Among all the 46 constraints, there are 6 first class constraints $\pi^{0}_{IJ}=0$
and 40 second class constraints. Therefore, the 2 local physical degrees of freedom remains.
All 24 torsion-free conditions appear as the consistency conditions for constraints.
Among the constraints, there are two scalar constraints (\eqref{2.9}, \eqref{2.10}) and one 2-dimensional vector constraint \eqref{2.11}.
The 6 Gauss constraints, \eqref{5.48}, \eqref{3.91}, \eqref{5.50} and \eqref{5.51}, are not independent.

The 4 Lagrange multipliers $n_{0},l_{0}$ and $e^{A}_{0}$ satisfy 8 differential equations \eqref{2.24}--\eqref{4.173}.
The integrability conditions of $n_{0},l_{0}$ and $e^{A}_{0}$ are Ricci identities.
The Lagrange multipliers, $\lambda^{A}_{a}$, $\lambda^{-A}_{1}$, $\lambda^{-+}_{1},\lambda^{-A}_{a},\lambda^{23}_{1},\lambda^{23}_{a}$
are completely solved (expressed by coframe and connections). The Lagrange multipliers $\lambda^{-+}_{a}$ and $\lambda^{+A}_{1}$ satisfy 2 algebraic equations and 2 differential equations. The Lagrange multipliers $\lambda^{+A}_{a}$ satisfy 1 algebraic equation
and 3 differential equations.

From the analysis, one will see that $\omega^{IJ}_{0}$ could also be treated as Lagrange multipliers,
because they are multiplied by the Gauss constraints. In this treatment, the Gauss constraints become primary constraints.
The consistency conditions containing $\omega^{IJ}_{0}$ are not treated as constraints but equations of multipliers.
The final degrees of freedom in phase space will be the same.

Using \eqref{3.3}, one can also replace $e^{A}_{a}$ by $\pi^{b}_{-B}$, so all the canonical variables in the Hamiltonian are $\omega^{IJ}_{\mu}$
and their conjugate momenta $\pi^{\mu}_{IJ}$. In this way, the dynamics of gravity is recovered as the pure connection dynamics.
However, the Hamiltonian analysis under this formalism will become more complicated.

The usual 1+3 spacelike foliation can be used in the initial-value analysis of the whole spacetime, while our foliation
can only be used in a small part of the whole spacetime within a short period of time. During this short period of time,
we can think there is just gravitational wave from one direction passing through a certain point in the spacetime.
In 1+3 foliation, there is 1 scalar constraint and a 3-dimensional vector constraint, while in our decomposition,
there are 2 scalar constraints and a 2-dimensional vector constraint. In the $\mathfrak{su}(2)$-connection dynamics,
there are 3 Gauss constraints corresponding to 3 generator of the $\mathfrak{su}(2)$ connection, as independent constraints,
but in our analysis, there are 6 Gauss constraints corresponding to 6 generators of the $\mathfrak{so}(1,3)(=\mathfrak{so}(1,1)\oplus\mathfrak{so}(2)
\oplus\mathfrak{t}^{-}(2)\oplus\mathfrak{t}^{+}(2))$ connection, which are not independent constraints. Besides, in $\mathfrak{su}(2)$-connection dynamics, frame rather than coframe is used, so the torsion-free conditions do not show up, while in our approach,
coframe is used, so the torsion-free conditions will show up as the requirements of consistency.
However, in all the formalisms, there are 2 local physical degrees of freedom.
The decomposition of symmetry and connection in the usual 1+3 way can not be postulated to higher dimensional spacetime,
while our decomposition can be applied to higher dimensions in principle.

The success of the Hamiltonian analysis of gravity in 3 and 4-dimensional spacetime shows that there will be probably no conceptual difficulty
for the Hamiltonian analysis of gravity in higher dimensional spacetime, but the analysis will become much harder technically.

\section*{Acknowledgment}

We would like to thank Zhe Chang, Yong-Chang Huang, Yi Ling, Yong-Ge Ma, Xiao-Ning Wu, Jing-Bo Wang and Bo-Feng Wu
for their helpful advice. Shi-Bei Kong would also like to thank Jun-Bao Wu, Yu Han, Fei Huang, Peng Liu and Xiang-Dong Zhang
for the good suggestions. This work is supported by National Natural Science Foundation of China under the grant 11275207.


\appendix

\section{The Proof of  $\eqref{2.62}$}

\begin{alignat}{1}
&\epsilon_{AB}\epsilon^{ab}(\Lambda^{-A}_{a,1}-\Lambda^{-+}_{1}\omega^{-A}_{a}-\omega^{-+}_{1}\Lambda^{-A}_{a}
-\Lambda^{-C}_{a}\omega^{DA}_{1}\delta_{CD})e^{B}_{b}+\epsilon_{AB}\epsilon^{ab}F^{-A}_{1a}\Lambda^{B}_b
\nonumber \\
=&(\epsilon_{AB}\epsilon^{ab}\Lambda^{-A}_{a}e^{B}_{b})_{,1}-\epsilon_{AB}\epsilon^{ab}\Lambda^{-A}_{a}e^{B}_{b,1}
-\epsilon_{AB}\epsilon^{ab}\Lambda^{-C}_{a}\omega^{DA}_{1}\delta_{CD}e^{B}_{b}-\epsilon_{AB}\epsilon^{ab}\omega^{-A}_{a}e^{B}_{b}\Lambda^{-+}_{1}
\nonumber \\&
-\epsilon_{AB}\epsilon^{ab}\Lambda^{-A}_{a}e^{B}_{b}\omega^{-+}_{1}+\epsilon_{AB}\epsilon^{ab}F^{-A}_{1a}\Lambda^{B}_b
\nonumber \\
\approx&(\epsilon_{AB}\epsilon^{ab}\Lambda^{-A}_{a}e^{B}_{b})_{,1}-\epsilon_{AB}\epsilon^{ab}\Lambda^{-A}_{a}\omega^{-B}_{b}
-\epsilon_{AB}\epsilon^{ab}\omega^{-A}_{a}e^{B}_{b}\Lambda^{-+}_{1}-\epsilon_{AB}\epsilon^{ab}\Lambda^{-A}_{a}e^{B}_{b}\omega^{-+}_{1}
+\epsilon_{AB}\epsilon^{ab}F^{-A}_{1a}\Lambda^{B}_b
\nonumber \\
=&(\epsilon_{AB}\epsilon^{ab}\Lambda^{-A}_{a}e^{B}_{b})_{,1}-\epsilon_{AB}\epsilon^{ab}\Lambda^{-A}_{a}e^{B}_{b}\omega^{-+}_{1}
-(\epsilon_{AB}\epsilon^{ab}\Lambda^{-B}_{b}+\epsilon_{AB}\epsilon^{ab}e^{B}_{b}\Lambda^{-+}_{1})\omega^{-A}_{a}
+\epsilon_{AB}\epsilon^{ab}F^{-A}_{1a}\Lambda^{B}_b
\nonumber \\
\approx&[\epsilon_{AB}\epsilon^{ab}(\omega^{-A}_{0,a}+\omega^{-+}_{0}\omega^{-A}_{a}-\omega^{-+}_{a}\omega^{-A}_{0}
-\omega^{-D}_{0}\omega^{CA}_{a}\delta_{DC}+\omega^{-D}_{a}\omega^{CA}_{0}\delta_{DC})e^B_b
-\epsilon_{AB}F^{-A}_{23}e^{B}_{0}-n_{0}F^{23}_{23}]_{,1}
\nonumber \\&
-\epsilon_{AB}\epsilon^{ab}(\omega^{-A}_{0,a}+\omega^{-+}_{0}\omega^{-A}_{a}-\omega^{-+}_{a}\omega^{-A}_{0}
-\omega^{-D}_{0}\omega^{CA}_{a}\delta_{DC}+\omega^{-D}_{a}\omega^{CA}_{0}\delta_{DC})e^B_b\omega^{-+}_{1}
+\epsilon_{AB}F^{-A}_{23}e^{B}_{0}\omega^{-+}_{1}
\nonumber \\&
+n_{0}F^{23}_{23}\omega^{-+}_{1}-[\epsilon_{AB}\epsilon^{ab} e^B_b(\omega^{-+}_{0,1}
+\omega^{-C}_{1}\omega^{D+}_{0}\delta_{CD}-\omega^{-C}_{0}\omega^{D+}_{1}\delta_{CD})
+\epsilon_{AB}\epsilon^{ab}(l_0F^{-B}_{1b}-e^{B}_0F^{-+}_{1b}-n_{0}F^{+B}_{1b})
\nonumber \\ &
+\epsilon_{AB}\epsilon^{ab}(\omega^{-B}_{0,b}+\omega^{-+}_{0}\omega^{-B}_{b}-\omega^{-+}_{b}\omega^{-B}_{0}
-\omega^{-D}_{0}\omega^{CB}_{b}\delta_{DC}+\omega^{-D}_{b}\omega^{CB}_{0}\delta_{DC})]\omega^{-A}_{a}
\nonumber \\ &
+\epsilon_{AB}\epsilon^{ab}(e^{A}_{0,a}+\omega^{-A}_{a}l_0+\omega^{+A}_{a}n_{0}+\omega^{AC}_{a}e^{D}_0\delta_{CD}
-\omega^{AC}_{0}e^D_a\delta_{CD})F^{-B}_{1b}
\nonumber \\
=&\epsilon_{AB}\epsilon^{ab}(\omega^{-A}_{0,a}+\omega^{-+}_{0}\omega^{-A}_{a}-\omega^{-+}_{a}\omega^{-A}_{0}
-\omega^{-D}_{0}\omega^{CA}_{a}\delta_{DC}+\omega^{-D}_{a}\omega^{CA}_{0}\delta_{DC})_{,1}e^B_b
\nonumber \\&
+\epsilon_{AB}\epsilon^{ab}(\omega^{-A}_{0,a}+\omega^{-+}_{0}\omega^{-A}_{a}-\omega^{-+}_{a}\omega^{-A}_{0}
-\omega^{-D}_{0}\omega^{CA}_{a}\delta_{DC}+\omega^{-D}_{a}\omega^{CA}_{0}\delta_{DC})e^B_{b,1}
\nonumber \\&
-\epsilon_{AB}\epsilon^{ab}(\omega^{-A}_{0,a}+\omega^{-+}_{0}\omega^{-A}_{a}-\omega^{-+}_{a}\omega^{-A}_{0}
-\omega^{-D}_{0}\omega^{CA}_{a}\delta_{DC}+\omega^{-D}_{a}\omega^{CA}_{0}\delta_{DC})e^B_b\omega^{-+}_{1}
\nonumber \\&
-\epsilon_{AB}\epsilon^{ab}(\omega^{-B}_{0,b}+\omega^{-+}_{0}\omega^{-B}_{b}-\omega^{-+}_{b}\omega^{-B}_{0}
-\omega^{-D}_{0}\omega^{CB}_{b}\delta_{DC}+\omega^{-D}_{b}\omega^{CB}_{0}\delta_{DC})\omega^{-A}_{a}
\nonumber \\ &
+\epsilon_{AB}\epsilon^{ab}e^{A}_{0,a}F^{-B}_{1b}+\epsilon_{AB}\epsilon^{ab}\omega^{-A}_{a}l_0F^{-B}_{1b}
+\epsilon_{AB}\epsilon^{ab}\omega^{+A}_{a}n_{0}F^{-B}_{1b}+\epsilon_{AB}\epsilon^{ab}\omega^{AC}_{a}e^{D}_0\delta_{CD}F^{-B}_{1b}
\nonumber \\&
-\epsilon_{AB}\epsilon^{ab} e^B_b(\omega^{-+}_{0,1}+\omega^{-C}_{1}\omega^{D+}_{0}\delta_{CD}-\omega^{-C}_{0}\omega^{D+}_{1}\delta_{CD})\omega^{-A}_{a}
-\epsilon_{AB}\epsilon^{ab}\omega^{AC}_{0}e^D_a\delta_{CD}F^{-B}_{1b}
\nonumber \\&
-\epsilon_{AB}F^{-A}_{23,1}e^{B}_{0}-\epsilon_{AB}F^{-A}_{23}e^{B}_{0,1}-n_{0,1}F^{23}_{23}-n_{0}F^{23}_{23,1}
+\epsilon_{AB}F^{-A}_{23}e^{B}_{0}\omega^{-+}_{1}+n_{0}F^{23}_{23}\omega^{-+}_{1}
\nonumber \\&
+\epsilon_{AB}\epsilon^{ab}e^{B}_0F^{-+}_{1b}\omega^{-A}_{a}+n_{0}\epsilon_{AB}\epsilon^{ab}F^{+B}_{1b}\omega^{-A}_{a}
-l_0\epsilon_{AB}\epsilon^{ab}F^{-B}_{1b}\omega^{-A}_{a}
\nonumber \\
\approx&\epsilon_{AB}\epsilon^{ab}(\omega^{-A}_{0,1})_{,a}e^B_b+\epsilon_{AB}\epsilon^{ab}\omega^{-+}_{0,1}\omega^{-A}_{a}e^B_b
+\epsilon_{AB}\epsilon^{ab}\omega^{-+}_{0}\omega^{-A}_{a,1}e^B_b-\epsilon_{AB}\epsilon^{ab}\omega^{-+}_{a,1}\omega^{-A}_{0}e^B_b
-\epsilon_{AB}\epsilon^{ab}\omega^{-+}_{a}\omega^{-A}_{0,1}e^B_b
\nonumber \\&
-\epsilon_{AB}\epsilon^{ab}\omega^{-D}_{0,1}\omega^{CA}_{a}\delta_{DC}e^B_b
-\epsilon_{AB}\epsilon^{ab}\omega^{-D}_{0}\omega^{CA}_{a,1}\delta_{DC}e^B_b
+\epsilon_{AB}\epsilon^{ab}\omega^{-D}_{a,1}\omega^{CA}_{0}\delta_{DC}e^B_b
+\epsilon_{AB}\epsilon^{ab}\omega^{-D}_{a}\omega^{CA}_{0,1}\delta_{DC}e^B_b
\nonumber \\&
-\epsilon_{AB}\epsilon^{ab}(\omega^{-A}_{0,a}+\omega^{-+}_{0}\omega^{-A}_{a}-\omega^{-+}_{a}\omega^{-A}_{0}
-\omega^{-D}_{0}\omega^{CA}_{a}\delta_{DC}+\omega^{-D}_{a}\omega^{CA}_{0}\delta_{DC})\omega^{BE}_{1}e^{F}_{b}\delta_{EF}
\nonumber \\&
-\epsilon_{AB}\epsilon^{ab}(\omega^{-A}_{0,a}+\omega^{-+}_{0}\omega^{-A}_{a}-\omega^{-+}_{a}\omega^{-A}_{0}
-\omega^{-D}_{0}\omega^{CA}_{a}\delta_{DC}+\omega^{-D}_{a}\omega^{CA}_{0}\delta_{DC})e^B_b\omega^{-+}_{1}
\nonumber \\&
-\epsilon_{AB}\epsilon^{ab} e^B_b(\omega^{-+}_{0,1}-\omega^{-C}_{0}\omega^{D+}_{1}\delta_{CD})\omega^{-A}_{a}
-n_{0}F^{23}_{23,1}+n_{0}\epsilon_{AB}\epsilon^{ab}\omega^{+A}_{a}F^{-B}_{1b}+n_{0}\epsilon_{AB}\epsilon^{ab}F^{+B}_{1b}\omega^{-A}_{a}
\nonumber \\ &
+n_{0}\epsilon_{AB}F^{-A}_{23}\omega^{+B}_{1}+\epsilon_{AB}\epsilon^{ab}\omega^{AC}_{a}e^{D}_0\delta_{CD}F^{-B}_{1b}
-\epsilon_{AB}F^{-A}_{23,1}e^{B}_{0}+\epsilon_{AB}F^{-A}_{23}e^{B}_{0}\omega^{-+}_{1}
+\epsilon_{AB}\epsilon^{ab}e^{B}_0F^{-+}_{1b}\omega^{-A}_{a}
\nonumber \\&
+\epsilon_{AB}F^{-A}_{23}\omega^{BC}_{1}e^{D}_{0}\delta_{CD}
+\omega^{-A}_{0}\epsilon_{AB}\epsilon^{ab}F^{-+}_{1a}e^{B}_{b}+\epsilon_{AB}\epsilon^{ab}e^{A}_{0,a}F^{-B}_{1b}
\nonumber\\
\approx&\omega^{-+}_{1,c}\omega^{-A}_{0}\epsilon_{AD}\epsilon^{cd}e^D_d+\omega^{-+}_{1}\omega^{-A}_{0,c}\epsilon_{AD}\epsilon^{cd}e^D_d
+\omega^{-B}_{0,c}\omega^{CA}_{1}\delta_{BC}\epsilon_{AD}\epsilon^{cd}e^D_d+\omega^{-B}_{0}\omega^{CA}_{1,c}\delta_{BC}\epsilon_{AD}\epsilon^{cd}e^D_d
\nonumber \\&
+(n_{0}\epsilon^{ab}e^{A}_{a}F^{23}_{1b}e^{-1})_{,c}\epsilon_{AD}\epsilon^{cd}e^D_d
+(\epsilon^{ab}e^{A}_{a}F^{-B}_{1b}e^{C}_0\epsilon_{BC}e^{-1})_{,c}\epsilon_{AD}\epsilon^{cd}e^D_d
-\epsilon_{AB}\epsilon^{ab}\omega^{-+}_{a}\omega^{-A}_{0,1}e^B_b
\nonumber \\&
-\epsilon_{AB}\epsilon^{ab}\omega^{-D}_{0,1}\omega^{CA}_{a}\delta_{DC}e^B_b
-\epsilon_{AB}\epsilon^{ab}\omega^{-A}_{0,a}\omega^{BE}_{1}e^{F}_{b}\delta_{EF}
-\epsilon_{AB}\epsilon^{ab}\omega^{-A}_{0,a}e^B_b\omega^{-+}_{1}
+\epsilon_{AB}\epsilon^{ab}\omega^{-+}_{0}\omega^{-A}_{a,1}e^B_b
\nonumber \\&
-\epsilon_{AB}\epsilon^{ab}\omega^{-+}_{a,1}\omega^{-A}_{0}e^B_b
-\epsilon_{AB}\epsilon^{ab}\omega^{-D}_{0}\omega^{CA}_{a,1}\delta_{DC}e^B_b
+\epsilon_{AB}\epsilon^{ab}\omega^{-D}_{a,1}\omega^{CA}_{0}\delta_{DC}e^B_b
+\epsilon_{AB}\epsilon^{ab}\omega^{-+}_{a}\omega^{-A}_{0}\omega^{BE}_{1}e^{F}_{b}\delta_{EF}
\nonumber \\&
+\epsilon_{AB}\epsilon^{ab}\omega^{-D}_{0}\omega^{CA}_{a}\delta_{DC}\omega^{BE}_{1}e^{F}_{b}\delta_{EF}
-\epsilon_{AB}\epsilon^{ab}\omega^{-D}_{a}\omega^{CA}_{0}\delta_{DC}\omega^{BE}_{1}e^{F}_{b}\delta_{EF}
-\epsilon_{AB}\epsilon^{ab}\omega^{-+}_{0}\omega^{-A}_{a}e^B_b\omega^{-+}_{1}
\nonumber \\&
+\epsilon_{AB}\epsilon^{ab}\omega^{-+}_{a}\omega^{-A}_{0}e^B_b\omega^{-+}_{1}
+\epsilon_{AB}\epsilon^{ab}\omega^{-D}_{0}\omega^{CA}_{a}\delta_{DC}e^B_b\omega^{-+}_{1}
+\epsilon_{AB}\epsilon^{ab} e^B_b\omega^{-C}_{0}\omega^{D+}_{1}\delta_{CD}\omega^{-A}_{a}
-n_{0}F^{23}_{23,1}
\nonumber \\ &
+n_{0}\epsilon_{AB}\epsilon^{ab}\omega^{+A}_{a}F^{-B}_{1b}+n_{0}\epsilon_{AB}\epsilon^{ab}F^{+B}_{1b}\omega^{-A}_{a}
+n_{0}\epsilon_{AB}F^{-A}_{23}\omega^{+B}_{1}-\epsilon_{AB}F^{-A}_{23,1}e^{B}_{0}
+\epsilon_{AB}\epsilon^{ab}\omega^{AC}_{a}e^{D}_0\delta_{CD}F^{-B}_{1b}
\nonumber \\&
+\epsilon_{AB}F^{-A}_{23}e^{B}_{0}\omega^{-+}_{1}+\epsilon_{AB}\epsilon^{ab}e^{B}_0F^{-+}_{1b}\omega^{-A}_{a}
+\epsilon_{AB}F^{-A}_{23}\omega^{BC}_{1}e^{D}_{0}\delta_{CD}+\omega^{-A}_{0}\epsilon_{AB}\epsilon^{ab}F^{-+}_{1a}e^{B}_{b}
+\epsilon_{AB}\epsilon^{ab}e^{A}_{0,a}F^{-B}_{1b}
\nonumber \\
=&n_{0}\epsilon^{ab}e_{,a}F^{23}_{1b}e^{-1}
+n_{0}\epsilon^{ab}F^{23}_{1b}e^{-1}_{,c}\epsilon^{cd}\epsilon_{AD}e^{A}_{a}e^D_d
+n_{0}\epsilon^{ab}F^{23}_{1b}e^{-1}\epsilon_{AD}\epsilon^{cd}e^{A}_{a}e^D_{c,d}
-n_{0}F^{23}_{23,1}
\nonumber \\&
+n_{0}\epsilon^{ab}F^{23}_{1b,c}e^{-1}\epsilon^{cd}\epsilon_{AD}e^{A}_{a}e^D_d
+n_{0}\epsilon_{AB}\epsilon^{ab}\omega^{+A}_{a}F^{-B}_{1b}
+n_{0}\epsilon_{AB}\epsilon^{ab}F^{+B}_{1b}\omega^{-A}_{a}+n_{0}\epsilon_{AB}F^{-A}_{23}\omega^{+B}_{1}
\nonumber \\&
-\epsilon_{DB}\epsilon^{cd}\omega^{ED}_{c}\delta_{AE}e^B_dn_{0}\epsilon^{ab}e^{A}_{a}F^{23}_{1b}e^{-1}
+\epsilon_{AB}\epsilon^{ab}\omega^{AC}_{a}e^{D}_0\delta_{CD}F^{-B}_{1b}
-\epsilon_{AB}F^{-A}_{23,1}e^{B}_{0}+\epsilon_{AB}F^{-A}_{23}e^{B}_{0}\omega^{-+}_{1}
\nonumber \\&
+\epsilon_{AB}\epsilon^{ab}e^{B}_0F^{-+}_{1b}\omega^{-A}_{a}+\epsilon_{AB}F^{-A}_{23}\omega^{BC}_{1}e^{D}_{0}\delta_{CD}
+\epsilon^{ab}e^{A}_{a,c}F^{-B}_{1b}e^{C}_0\epsilon_{BC}e^{-1}\epsilon_{AD}\epsilon^{cd}e^D_d
\nonumber \\&
+\epsilon^{ab}e^{A}_{a}F^{-B}_{1b,c}e^{C}_0\epsilon_{BC}e^{-1}\epsilon_{AD}\epsilon^{cd}e^D_d
-\omega^{-C}_{c}e^{D}_{0}\delta_{CD}\epsilon^{ab}e^{A}_{a}F^{23}_{1b}e^{-1}\epsilon_{AB}\epsilon^{cd}e^B_d
+\epsilon^{ab}e^{A}_{a}F^{-B}_{1b}e^{C}_0\epsilon_{BC}e^{-1}_{,c}\epsilon_{AD}\epsilon^{cd}e^D_d
\nonumber \\&
-\epsilon_{AD}\epsilon^{cd}\omega^{-+}_{c}e^D_d\epsilon^{ab}e^{A}_{a}F^{-B}_{1b}e^{C}_0\epsilon_{BC}e^{-1}
-\epsilon_{DF}\epsilon^{cd}\omega^{ED}_{c}\delta_{AE}e^F_d\epsilon^{ab}e^{A}_{a}F^{-B}_{1b}e^{C}_0\epsilon_{BC}e^{-1}
\nonumber
\end{alignat}
\begin{alignat}{1}
\approx&-n_{0}\epsilon^{ab}F^{23}_{1a,b}-n_{0}F^{23}_{23,1}+n_{0}\epsilon_{AB}\epsilon^{ab}\omega^{+A}_{a}F^{-B}_{1b}
+n_{0}\epsilon_{AB}\epsilon^{ab}F^{+B}_{1b}\omega^{-A}_{a}+n_{0}\epsilon_{AB}F^{-A}_{23}\omega^{+B}_{1}
\nonumber \\&
+e^{C}_0\epsilon_{AB}\epsilon^{ab}F^{-B}_{1b}\omega^{AD}_{a}\delta_{CD}
-\epsilon_{AB}F^{-A}_{23,1}e^{B}_{0}+\epsilon_{AB}F^{-A}_{23}e^{B}_{0}\omega^{-+}_{1}
+\epsilon_{AB}F^{-A}_{23}\omega^{BC}_{1}e^{D}_{0}\delta_{CD}
\nonumber \\&
+e^{A}_{0}\epsilon_{AB}\epsilon^{ab}F^{-B}_{1a,b}
+e^{B}_0\epsilon_{AB}\epsilon^{ab}F^{-+}_{1b}\omega^{-A}_{a}
-e^{B}_0\delta_{AB}\epsilon^{ab}F^{23}_{1b}\omega^{-A}_{a}
+e^{A}_0\epsilon_{AB}\epsilon^{ab}\omega^{-+}_{a}F^{-B}_{1b}
\nonumber \\
\approx&n_{0}\epsilon_{AB}F^{-A}_{23}\omega^{+B}_{1}
+n_{0}\epsilon^{ab}\epsilon_{AB}\omega^{-A}_{a,b}\omega^{+B}_{1}
+n_{0}\epsilon_{AB}\epsilon^{ab}F^{-A}_{1a}\omega^{+B}_{b}
-n_{0}\epsilon^{ab}\epsilon_{AB}\omega^{-A}_{a,1}\omega^{+B}_{b}
+n_{0}\epsilon^{ab}\epsilon_{AB}\omega^{-A}_{a}\omega^{+B}_{1,b}
\nonumber \\&
-n_{0}\epsilon^{ab}\epsilon_{AB}\omega^{-A}_{a}\omega^{+B}_{b,1}
+n_{0}\epsilon_{AB}\epsilon^{ab}F^{+B}_{1b}\omega^{-A}_{a}
+e^{B}_0\epsilon_{AB}\epsilon^{ab}F^{-+}_{1b}\omega^{-A}_{a}
-\epsilon_{AB}e^{B}_{0}\epsilon^{ab}\omega^{-+}_{b,1}\omega^{-A}_{a}
\nonumber \\&
+e^{B}_{0}\epsilon_{AB}\epsilon^{ab}\omega^{-+}_{1,b}\omega^{-A}_{a}
+e^{A}_0\epsilon_{AB}\epsilon^{ab}\omega^{-+}_{a}F^{-B}_{1b}
-\epsilon_{AB}e^{A}_{0}\epsilon^{ab}\omega^{-+}_{a}\omega^{-B}_{b,1}
+e^{C}_0\epsilon_{AB}\epsilon^{ab}F^{-B}_{1b}\omega^{AD}_{a}\delta_{CD}
\nonumber \\&
-\epsilon_{AC}e^{C}_{0}\epsilon^{ab}\omega^{-B}_{b,1}\omega^{AD}_{a}\delta_{BD}
-e^{B}_0\delta_{AB}\epsilon^{ab}F^{23}_{1b}\omega^{-A}_{a}
-e^{B}_{0}\epsilon^{ab}\omega^{-A}_{a}\omega^{23}_{1,b}\delta_{AB}
+e^{B}_{0}\epsilon^{ab}\omega^{-A}_{a}\omega^{23}_{b,1}\delta_{AB}
\nonumber \\&
+e^{B}_{0}\epsilon_{AB}\epsilon^{ab}\omega^{-+}_{1}\omega^{-A}_{a,b}
+\epsilon_{AB}F^{-A}_{23}e^{B}_{0}\omega^{-+}_{1}
-e^{B}_{0}\epsilon^{ab}\omega^{-A}_{a,b}\omega^{23}_{1}\delta_{AB}
-e^{B}_{0}F^{-A}_{23}\omega^{23}_{1}\delta_{AB}
\nonumber \\
\approx&n_{0}\epsilon_{AB}F^{-A}_{23}\omega^{+B}_{1}
-n_{0}\epsilon_{AB}F^{-A}_{23}\omega^{+B}_{1}
+n_{0}\epsilon_{AB}\epsilon^{ab}F^{-A}_{1a}\omega^{+B}_{b}
-n_{0}\epsilon^{ab}\epsilon_{AB}F^{-A}_{1a}\omega^{+B}_{b}
-n_{0}\epsilon^{ab}\epsilon_{AB}\omega^{-A}_{a}F^{+B}_{1b}
\nonumber \\&
+n_{0}\epsilon_{AB}\epsilon^{ab}F^{+B}_{1b}\omega^{-A}_{a}
+e^{B}_0\epsilon_{AB}\epsilon^{ab}F^{-+}_{1b}\omega^{-A}_{a}
-\epsilon_{AB}e^{B}_{0}\epsilon^{ab}F^{-+}_{1b}\omega^{-A}_{a}
+e^{A}_0\epsilon_{AB}\epsilon^{ab}\omega^{-+}_{a}F^{-B}_{1b}
\nonumber \\&
-\epsilon_{AB}e^{A}_{0}\epsilon^{ab}\omega^{-+}_{a}F^{-B}_{1b}
+e^{C}_0\epsilon_{AB}\epsilon^{ab}F^{-B}_{1b}\omega^{AD}_{a}\delta_{CD}
-\epsilon_{AC}e^{C}_{0}\epsilon^{ab}F^{-B}_{1b}\omega^{AD}_{a}\delta_{BD}
-e^{B}_0\delta_{AB}\epsilon^{ab}F^{23}_{1b}\omega^{-A}_{a}
\nonumber \\&
+e^{B}_{0}\epsilon^{ab}\omega^{-A}_{a}F^{23}_{1b}\delta_{AB}
-e^{B}_{0}\epsilon_{AB}\epsilon^{ab}\omega^{-+}_{1}F^{-A}_{ab}
+\epsilon_{AB}F^{-A}_{23}e^{B}_{0}\omega^{-+}_{1}
+e^{B}_{0}\epsilon^{ab}F^{-A}_{ab}\omega^{23}_{1}\delta_{AB}
-e^{B}_{0}F^{-A}_{23}\omega^{23}_{1}\delta_{AB}=0. \nonumber
\end{alignat}
In the 1st ``$\approx$", \eqref{4.173} have been used. While in the 2nd ``$\approx$", \eqref{2.9} and \eqref{2.10} have been used.
The identity $\epsilon^{ab}F^{-A}_{1a}e^{B}_{b}\delta_{AB}\approx0$ and \eqref{2.26}, \eqref{4.173}, \eqref{5.158}, \eqref{2.31-1}
have been used in the 3rd ``$\approx$", and \eqref{2.11}, \eqref{2.26}, \eqref{2.27}, \eqref{4.173}, \eqref{5.158}, \eqref{2.31-1}, \eqref{5.162}
have been used in the 4th ``$\approx$". In the 5th ``$\approx$", \eqref{2.33} have been used.
\eqref{2.31-1} has been used in the 6th and last ``$\approx$".

The proof of the additional identity:
\begin{alignat}{1}
\epsilon^{ab}F^{-A}_{1a}e^{B}_{b}\delta_{AB}
=&\epsilon^{ab}(\omega^{-A}_{a,1}-\omega^{-A}_{1,a}-\omega^{-+}_{1}\omega^{-A}_{a}+\omega^{-C}_{1}\omega^{DA}_{a}\delta_{CD}
-\omega^{-C}_{a}\omega^{DA}_{1}\delta_{CD}+\omega^{-+}_{a}\omega^{-A}_{1})e^{B}_{b}\delta_{AB}
\nonumber \\
\approx&\epsilon^{ab}(\omega^{-A}_{a,1}-\omega^{-C}_{a}\omega^{DA}_{1}\delta_{CD})e^{B}_{b}\delta_{AB}
\nonumber \\
=&\delta_{AB}\epsilon^{ab}\omega^{-A}_{a,1}e^{B}_{b}-\delta_{AB}\epsilon^{ab}\omega^{-C}_{a}\omega^{DA}_{1}\delta_{CD}e^{B}_{b}
\nonumber \\
=&(\delta_{AB}\epsilon^{ab}\omega^{-A}_{a}e^{B}_{b})_{,1}-\delta_{AB}\epsilon^{ab}\omega^{-A}_{a}e^{B}_{b,1}
-\delta_{AB}\epsilon^{ab}\omega^{-C}_{a}\omega^{DA}_{1}\delta_{CD}e^{B}_{b}
\nonumber \\
\approx&-\delta_{CD}\epsilon^{ab}\omega^{-C}_{a}e^{D}_{b,1}
-\delta_{AB}\epsilon^{ab}\omega^{-C}_{a}\omega^{DA}_{1}\delta_{CD}e^{B}_{b}
\nonumber \\
=&-\delta_{CD}\epsilon^{ab}\omega^{-C}_{a}(e^{D}_{b,1}+\omega^{DA}_{1}e^{B}_{b}\delta_{AB})
\nonumber \\
\approx&-\delta_{CD}\epsilon^{ab}\omega^{-C}_{a}\omega^{-D}_{b}=0.\nonumber
\end{alignat}
In the 1st ``$\approx$", \eqref{2.31-1} and \eqref{5.162} have been used.
In the 2nd and 3rd ``$\approx$", \eqref{5.158} and \eqref{5.162} have been used respectively.

\section{The Proof of $\eqref{2.47}$}

\begin{alignat}{1}
&\epsilon^{ab}(\omega^{-A}_{0,a}-\omega^{-+}_{a}\omega^{-A}_{0}-\omega^{-B}_{0}\omega^{CA}_{a}\delta_{BC}
+\omega^{-B}_{a}\omega^{CA}_{0}\delta_{BC})e^D_b\delta_{AD}
+\epsilon^{ab}(l_0F^{-A}_{1a}-e^{A}_0F^{-+}_{1a}-n_{0}F^{+A}_{1a})e^B_b\delta_{AB}
\nonumber \\ &
-\epsilon^{ab}(e^{A}_{0,a}+\omega^{-A}_{a}l_0+\omega^{+A}_{a}n_{0}+\omega^{AB}_{a}e^{C}_0\delta_{BC}-\omega^{AB}_{0}e^C_a\delta_{BC})
\omega^{-D}_{b}\delta_{AD}
\nonumber \\
\approx&-\epsilon^{ab}e^{A}_0F^{-+}_{1a}e^E_b\delta_{AE}-\epsilon^{ab}n_{0}F^{+A}_{1a}e^E_b\delta_{AE}
+\epsilon^{ab}(\omega^{-A}_{0,a}-\omega^{-+}_{a}\omega^{-A}_{0}-\omega^{-D}_{0}\omega^{CA}_{a}\delta_{DC})e^E_b\delta_{AE}
\nonumber \\ &
+\epsilon^{ab}\omega^{-A}_{a}(e^{B}_{0,b}+\omega^{+B}_{b}n_{0}+\omega^{BC}_{b}e^{D}_0\delta_{CD})\delta_{AB}
\nonumber \\
=&\epsilon^{ab}e^E_b\delta_{AE}\omega^{-A}_{0,a}-\epsilon^{ab}e^E_b\delta_{AE}\omega^{-+}_{a}\omega^{-A}_{0}
-\epsilon^{ab}e^E_b\delta_{AE}\omega^{-D}_{0}\omega^{CA}_{a}\delta_{DC}+\epsilon^{ab}\omega^{-A}_{a}\delta_{AB}e^{B}_{0,b}
+\epsilon^{ab}\omega^{-A}_{a}\delta_{AB}\omega^{+B}_{b}n_{0}
\nonumber \\ &
+\epsilon^{ab}\omega^{-A}_{a}\delta_{AB}\omega^{BC}_{b}e^{D}_0\delta_{CD}
-\epsilon^{ab}e^{A}_0F^{-+}_{1a}e^E_b\delta_{AE}-\epsilon^{ab}n_{0}F^{+A}_{1a}e^E_b\delta_{AE}
\nonumber \\
=&\epsilon^{ab}(\delta_{AE}\omega^{-A}_{0}e^E_b)_{,a}-\epsilon^{ab}\delta_{AE}\omega^{-A}_{0}e^E_{b,a}
-\epsilon^{ab}e^E_b\delta_{AE}\omega^{-+}_{a}\omega^{-A}_{0}-\epsilon^{ab}e^E_b\delta_{AE}\omega^{-D}_{0}\omega^{CA}_{a}\delta_{DC}
+\epsilon^{ab}(\omega^{-A}_{a}\delta_{AB}e^{B}_{0})_{,b}
\nonumber \\ &
-\epsilon^{ab}\omega^{-A}_{a,b}\delta_{AB}e^{B}_{0}+\epsilon^{ab}\omega^{-A}_{a}\delta_{AB}\omega^{+B}_{b}n_{0}
+\epsilon^{ab}\omega^{-A}_{a}\delta_{AB}\omega^{BC}_{b}e^{D}_0\delta_{CD}-\epsilon^{ab}e^{A}_0F^{-+}_{1a}e^E_b\delta_{AE}
-\epsilon^{ab}n_{0}F^{+A}_{1a}e^E_b\delta_{AE}
\nonumber \\
=&-\epsilon^{ab}(\delta_{AE}\omega^{-A}_{0}e^E_a)_{,b}+\epsilon^{ab}(\omega^{-A}_{a}\delta_{AB}e^{B}_{0})_{,b}
-\epsilon^{ab}\delta_{AE}\omega^{-A}_{0}e^E_{b,a}-\epsilon^{ab}e^E_b\delta_{AE}\omega^{-+}_{a}\omega^{-A}_{0}
-\epsilon^{ab}e^E_b\delta_{AE}\omega^{-D}_{0}\omega^{CA}_{a}\delta_{DC}
\nonumber \\ &
+\epsilon^{ab}\omega^{-A}_{a}\delta_{AB}\omega^{+B}_{b}n_{0}-\epsilon^{ab}n_{0}F^{+A}_{1a}e^E_b\delta_{AE}
-\epsilon^{ab}\omega^{-A}_{a,b}\delta_{AB}e^{B}_{0}+\epsilon^{ab}\omega^{-A}_{a}\delta_{AB}\omega^{BC}_{b}e^{D}_0\delta_{CD}
-\epsilon^{ab}e^{A}_0F^{-+}_{1a}e^E_b\delta_{AE}
\nonumber \\
\approx&\epsilon^{ab}(-n_{0,a}+\omega^{-+}_{a}n_{0})_{,b}+\epsilon^{ab}\delta_{AE}\omega^{-A}_{0}\omega^{EF}_{a}e^{G}_{b}\delta_{FG}
-\epsilon^{ab}e^E_b\delta_{AE}\omega^{-+}_{a}\omega^{-A}_{0}-\epsilon^{ab}e^E_b\delta_{AE}\omega^{-D}_{0}\omega^{CA}_{a}\delta_{DC}
\nonumber \\ &
+\epsilon^{ab}\omega^{-A}_{a}\delta_{AB}\omega^{+B}_{b}n_{0}-\epsilon^{ab}n_{0}F^{+A}_{1a}e^E_b\delta_{AE}
-\epsilon^{ab}\omega^{-A}_{a,b}\delta_{AB}e^{B}_{0}+\epsilon^{ab}\omega^{-A}_{a}\delta_{AB}\omega^{BC}_{b}e^{D}_0\delta_{CD}
-\epsilon^{ab}e^{A}_0F^{-+}_{1a}e^E_b\delta_{AE}\nonumber
\end{alignat}
\begin{alignat}{1}
=&\epsilon^{ab}(\omega^{-+}_{a,b}n_{0}+\omega^{-+}_{a}n_{0,b})
-\epsilon^{ab}e^E_b\delta_{AE}\omega^{-+}_{a}\omega^{-A}_{0}
+\epsilon^{ab}\omega^{-A}_{a}\delta_{AB}\omega^{+B}_{b}n_{0}
-\epsilon^{ab}n_{0}F^{+A}_{1a}e^E_b\delta_{AE}
\nonumber \\ &
-\epsilon^{ab}\omega^{-A}_{a,b}\delta_{AB}e^{B}_{0}
+\epsilon^{ab}\omega^{-A}_{a}\delta_{AB}\omega^{BC}_{b}e^{D}_0\delta_{CD}
-\epsilon^{ab}e^{A}_0F^{-+}_{1a}e^E_b\delta_{AE}
\nonumber \\
\approx&\epsilon^{ab}\omega^{-+}_{a,b}n_{0}-\epsilon^{ab}\omega^{-+}_{b}
(\omega^{-+}_{a}n_{0}+\omega^{-A}_{0}e^{B}_{a}\delta_{AB}-\omega^{-A}_{a}e^{B}_{0}\delta_{AB})
-\epsilon^{ab}e^E_b\delta_{AE}\omega^{-+}_{a}\omega^{-A}_{0}
+\epsilon^{ab}\omega^{-A}_{a}\delta_{AB}\omega^{+B}_{b}n_{0}
\nonumber \\ &
-\epsilon^{ab}n_{0}F^{+A}_{1a}e^E_b\delta_{AE}-\epsilon^{ab}\omega^{-A}_{a,b}\delta_{AB}e^{B}_{0}
+\epsilon^{ab}\omega^{-A}_{a}\delta_{AB}\omega^{BC}_{b}e^{D}_0\delta_{CD}-\epsilon^{ab}e^{A}_0F^{-+}_{1a}e^E_b\delta_{AE}
\nonumber \\
=&\epsilon^{ab}\omega^{-+}_{a,b}n_{0}+\epsilon^{ab}\omega^{-A}_{a}\delta_{AB}\omega^{+B}_{b}n_{0}
-\epsilon^{ab}n_{0}F^{+A}_{1a}e^E_b\delta_{AE}-\epsilon^{ab}\omega^{-A}_{a,b}\delta_{AB}e^{B}_{0}
+\epsilon^{ab}\omega^{-A}_{a}\delta_{AB}\omega^{BC}_{b}e^{D}_0\delta_{CD}
\nonumber \\ &
+\epsilon^{ab}\omega^{-+}_{b}\omega^{-A}_{a}e^{B}_{0}\delta_{AB}
-\epsilon^{ab}e^{A}_0F^{-+}_{1a}e^E_b\delta_{AE}
\nonumber \\
=&-n_{0}(F^{-+}_{23}+\epsilon^{ab}F^{+A}_{1a}e^E_b\delta_{AE})
+e^{A}_{0}\delta_{AB}(F^{-B}_{23}-\epsilon^{ab}F^{-+}_{1a}e^B_b)
\nonumber \\
\approx&-n_{0}(F^{-+}_{23}+\epsilon^{ab}F^{+A}_{1a}e^B_b\delta_{AB})
\nonumber \\
=&-n_{0}[\epsilon^{ab}(\omega^{+A}_{a,1}-\omega^{+A}_{1,a}+\omega^{-+}_{1}\omega^{+A}_{a}+\omega^{+C}_{1}\omega^{DA}_{a}\delta_{CD}
-\omega^{-+}_{a}\omega^{+A}_{1}-\omega^{+C}_{a}\omega^{DA}_{1}\delta_{CD})e^B_b\delta_{AB}
\nonumber \\&
+(\omega^{-+}_{3,2}-\omega^{-+}_{2,3}-\omega^{-A}_{2}\omega^{+B}_{3}\delta_{AB}+\omega^{-A}_{3}\omega^{+B}_{2}\delta_{AB})]
\nonumber \\
=&-n_{0}[(\omega^{-+}_{3,2}-\omega^{-+}_{2,3}-\omega^{-A}_{2}\omega^{+B}_{3}\delta_{AB}+\omega^{-A}_{3}\omega^{+B}_{2}\delta_{AB}
+(\epsilon^{ab}\omega^{+A}_{a}e^B_b\delta_{AB})_{,1}-\epsilon^{ab}\omega^{+A}_{a}e^B_{b,1}\delta_{AB}
\nonumber \\&
-(\epsilon^{ab}\omega^{+A}_{1}e^B_b\delta_{AB})_{,a}+\epsilon^{ab}\omega^{+A}_{1}e^B_{b,a}\delta_{AB}
+\epsilon^{ab}\omega^{-+}_{1}\omega^{+A}_{a}e^B_b\delta_{AB}
+\epsilon^{ab}\omega^{+C}_{1}\omega^{DA}_{a}\delta_{CD}e^B_b\delta_{AB}
\nonumber \\&
-\epsilon^{ab}\omega^{-+}_{a}\omega^{+A}_{1}e^B_b\delta_{AB}
-\epsilon^{ab}\omega^{+C}_{a}\omega^{DA}_{1}\delta_{CD}e^B_b\delta_{AB}]
\nonumber \\
\approx&-n_{0}(\epsilon^{ab}\omega^{+A}_{1}\omega^{BC}_{b}e^{D}_{a}\delta_{AB}
+\epsilon^{ab}\omega^{+C}_{1}\omega^{DA}_{a}\delta_{CD}e^B_b\delta_{AB})=0. \nonumber
\end{alignat}
In the 1st ``$\approx$", one identity $\epsilon^{ab}F^{-A}_{1a}e^{B}_{b}\delta_{AB}\approx0$ has been used.
While in the 2nd ``$\approx$", \eqref{2.27} and \eqref{2.33} have been used.
In the 3rd and 4th ``$\approx$", \eqref{2.31} and \eqref{2.11} have been used respectively.
In the last ``$\approx$", \eqref{2.31}, \eqref{2.25}, \eqref{2.26} and \eqref{2.27} have been used.

\section{Integrability of $n_{0}$}

The integrability of $n_{0}$ requires that
\begin{alignat}{1}
n_{0,1a}-n_{0,a1}=&0,\label{4.1}\\
\epsilon^{ab}n_{0,ab}=&0.\label{4.3}
\end{alignat}
From \eqref{2.24}, one can get
\begin{alignat}{1}
(n_{0,1}-\omega^{-+}_{1}n_{0})_{,a}
\approx&n_{0,1a}-\omega^{-+}_{1,a}n_{0}-\omega^{-+}_{1}n_{0,a}
\nonumber \\
\approx&n_{0,a1}-\omega^{-+}_{1,a}n_{0}-\omega^{-+}_{1}(\omega^{-+}_{a}n_{0}+\omega^{-A}_{0}e^{B}_{a}\delta_{AB}-\omega^{-A}_{a}e^{B}_{0}\delta_{AB})
\approx0.
\end{alignat}
On the other hand, from \eqref{2.31}, one can get
\begin{alignat}{1}
&(n_{0,a}-\omega^{-+}_{a}n_{0}-\omega^{-A}_{0}e^{B}_{a}\delta_{AB}+\omega^{-A}_{a}e^{B}_{0}\delta_{AB})_{,1}
\nonumber \\
\approx&n_{0,a1}-\omega^{-+}_{a,1}n_{0}-\omega^{-+}_{a}n_{0,1}-\omega^{-A}_{0,1}e^{B}_{a}\delta_{AB}-\omega^{-A}_{0}e^{B}_{a,1}\delta_{AB}
+\omega^{-A}_{a,1}e^{B}_{0}\delta_{AB}+\omega^{-A}_{a}e^{B}_{0,1}\delta_{AB}
\nonumber \\
\approx&n_{0,a1}-\omega^{-+}_{a,1}n_{0}-\omega^{-+}_{a}\omega^{-+}_{1}n_{0}-\omega^{-A}_{0,1}e^{B}_{a}\delta_{AB}
-\omega^{-B}_{0}(\omega^{-A}_{a}-\omega^{AC}_{1}e^{D}_{a}\delta_{CD})\delta_{AB}
+\omega^{-A}_{a,1}e^{B}_{0}\delta_{AB}
\nonumber \\&
+\omega^{-A}_{a}\delta_{AB}(\omega^{-B}_{0}-\omega^{+B}_{1}n_{0}-\omega^{BC}_{1}e^{D}_{0}\delta_{CD})
\nonumber \\
\approx&n_{0,a1}-\omega^{-+}_{a,1}n_{0}-\omega^{-+}_{a}\omega^{-+}_{1}n_{0}-\omega^{-A}_{a}\delta_{AB}\omega^{+B}_{1}n_{0}
-\omega^{-A}_{0,1}e^{B}_{a}\delta_{AB}+\omega^{-B}_{0}\omega^{AC}_{1}e^{D}_{a}\delta_{CD}\delta_{AB}
\nonumber \\&
+\omega^{-A}_{a,1}e^{B}_{0}\delta_{AB}-\omega^{-A}_{a}\delta_{AB}\omega^{BC}_{1}e^{D}_{0}\delta_{CD}\approx0.
\end{alignat}
\eqref{4.1} requires
\begin{alignat}{1}
&-\omega^{-+}_{1,a}n_{0}-\omega^{-+}_{1}\omega^{-+}_{a}n_{0}-\omega^{-+}_{1}\omega^{-A}_{0}e^{B}_{a}\delta_{AB}
+\omega^{-+}_{1}\omega^{-A}_{a}e^{B}_{0}\delta_{AB}
\nonumber \\
\approx&-\omega^{-+}_{a,1}n_{0}-\omega^{-+}_{a}\omega^{-+}_{1}n_{0}-\omega^{-A}_{a}\delta_{AB}\omega^{+B}_{1}n_{0}
-\omega^{-A}_{0,1}e^{B}_{a}\delta_{AB}+\omega^{-B}_{0}\omega^{AC}_{1}e^{D}_{a}\delta_{CD}\delta_{AB}
\nonumber \\&
+\omega^{-A}_{a,1}e^{B}_{0}\delta_{AB}-\omega^{-A}_{a}\delta_{AB}\omega^{BC}_{1}e^{D}_{0}\delta_{CD},
\end{alignat}
which is equivalent to
\begin{alignat}{1}
&\omega^{-+}_{a,1}n_{0}-\omega^{-+}_{1,a}n_{0}+\omega^{-A}_{a}\delta_{AB}\omega^{+B}_{1}n_{0}-\omega^{-+}_{1}\omega^{-A}_{0}e^{B}_{a}\delta_{AB}
+\omega^{-A}_{0,1}e^{B}_{a}\delta_{AB}-\omega^{-B}_{0}\omega^{AC}_{1}e^{D}_{a}\delta_{CD}\delta_{AB}
\nonumber \\&
+\omega^{-+}_{1}\omega^{-A}_{a}e^{B}_{0}\delta_{AB}-\omega^{-A}_{a,1}e^{B}_{0}\delta_{AB}
+\omega^{-A}_{a}\delta_{AB}\omega^{BC}_{1}e^{D}_{0}\delta_{CD}
\nonumber \\
\approx&F^{-+}_{1a}n_{0}+(\omega^{-A}_{0,1}-\omega^{-+}_{1}\omega^{-A}_{0}+\omega^{-C}_{0}\omega^{DA}_{1}\delta_{CD})e^{B}_{a}\delta_{AB}
-F^{-A}_{1a}e^{B}_{0}\delta_{AB}
\nonumber \\
\approx&F^{-+}_{1a}n_{0}-F^{-A}_{01}e^{B}_{a}\delta_{AB}-F^{-A}_{1a}e^{B}_{0}\delta_{AB}=-(\eta_{IJ}F^{-I}\wedge e^{J})_{01a}=0.\label{4.6}
\end{alignat}
The integrability conditions \eqref{4.6} are Ricci identities.

From \eqref{2.31}, one gets
\begin{alignat}{1}
\epsilon^{ab}n_{0,ab}\approx\epsilon^{ab}(\omega^{-+}_{a,b}n_{0}+\omega^{-+}_{a}n_{0,b}+\omega^{-B}_{0,b}e^{A}_{a}\delta_{AB}
+\omega^{-B}_{0}e^{A}_{a,b}\delta_{AB}-\omega^{-A}_{a,b}e^{B}_{0}\delta_{AB}-\omega^{-A}_{a}e^{B}_{0,b}\delta_{AB}).
\end{alignat}
\eqref{4.3} requires that
\begin{alignat}{1}
&\epsilon^{ab}(\omega^{-+}_{a,b}n_{0}+\omega^{-+}_{a}n_{0,b}+\omega^{-B}_{0,b}e^{A}_{a}\delta_{AB}+\omega^{-B}_{0}e^{A}_{a,b}\delta_{AB}
-\omega^{-A}_{a,b}e^{B}_{0}\delta_{AB}-\omega^{-A}_{a}e^{B}_{0,b}\delta_{AB})
\nonumber \\
\approx&\epsilon^{ab}[\omega^{-+}_{a,b}n_{0}+\omega^{-+}_{a}(\omega^{-+}_{b}n_{0}+\omega^{-A}_{0}e^{B}_{b}\delta_{AB}
-\omega^{-A}_{b}e^{B}_{0}\delta_{AB})+\omega^{-B}_{0,b}e^{A}_{a}\delta_{AB}+\omega^{-B}_{0}e^{A}_{a,b}\delta_{AB}
-\omega^{-A}_{a,b}e^{B}_{0}\delta_{AB}-\omega^{-A}_{a}e^{B}_{0,b}\delta_{AB}]
\nonumber \\
\approx&\epsilon^{ab}\omega^{-+}_{a,b}n_{0}+\epsilon^{ab}(-\omega^{-A}_{a,b}-\omega^{-+}_{a}\omega^{-A}_{b})e^{B}_{0}\delta_{AB}
-\epsilon^{ab}\omega^{-A}_{a}e^{B}_{0,b}\delta_{AB}+\epsilon^{ab}(e^{B}_{a,b}+\omega^{-+}_{a}e^{B}_{b})\omega^{-A}_{0}\delta_{AB}
-\epsilon^{ab}\omega^{-A}_{0,a}e^{B}_{b}\delta_{AB}
\nonumber \\
\approx&\epsilon^{ab}\omega^{-+}_{a,b}n_{0}+\epsilon^{ab}(-\omega^{-A}_{a,b}-\omega^{-+}_{a}\omega^{-A}_{b})e^{B}_{0}\delta_{AB}
-\epsilon^{ab}\omega^{-A}_{a}e^{B}_{0,b}\delta_{AB}+\epsilon^{ab}(\omega^{BC}_{a}e^{D}_{b}\delta_{CD}
+\omega^{-+}_{a}e^{B}_{b})\omega^{-A}_{0}\delta_{AB}-\epsilon^{ab}\omega^{-A}_{0,a}e^{B}_{b}\delta_{AB}
\nonumber \\
\approx&\epsilon^{ab}\omega^{-+}_{a,b}n_{0}+\epsilon^{ab}(-\omega^{-A}_{a,b}-\omega^{-+}_{a}\omega^{-A}_{b})e^{B}_{0}\delta_{AB}
-\epsilon^{ab}\omega^{-A}_{a}e^{B}_{0,b}\delta_{AB}
-\epsilon^{ab}(\omega^{-A}_{0,a}-\omega^{-+}_{a}\omega^{-A}_{0}-\omega^{-C}_{0}\omega^{DA}_{a}\delta_{CD})e^{B}_{b}\delta_{AB}
\nonumber \\
\approx&\epsilon^{ab}(\omega^{-+}_{a,b}-\omega^{+A}_{a}\omega^{-B}_{b}\delta_{AB})n_{0}
+\epsilon^{ab}(-\omega^{-A}_{a,b}-\omega^{-+}_{a}\omega^{-A}_{b}+\omega^{-C}_{a}\omega^{DA}_{b}\delta_{CD})e^{B}_{0}\delta_{AB}
\nonumber \\&
-\epsilon^{ab}\omega^{-A}_{a}(e^{B}_{0,b}+\omega^{-B}_{b}l_0+\omega^{+B}_{b}n_{0}+\omega^{BC}_{b}e^{D}_0\delta_{CD}
-\omega^{BC}_{0}e^D_b\delta_{CD})\delta_{AB}
\nonumber \\&
-\epsilon^{ab}(\omega^{-A}_{0,a}-\omega^{-+}_{a}\omega^{-A}_{0}-\omega^{-C}_{0}\omega^{DA}_{a}\delta_{CD}
+\omega^{-C}_{a}\omega^{DA}_{0}\delta_{CD}+\omega^{-+}_{0}\omega^{-A}_{a})e^{B}_{b}\delta_{AB}
\nonumber \\
\approx&-\epsilon^{ab}(\omega^{-A}_{0,a}-\omega^{-+}_{a}\omega^{-A}_{0}-\omega^{-C}_{0}\omega^{DA}_{a}\delta_{CD}
+\omega^{-C}_{a}\omega^{DA}_{0}\delta_{CD}+\omega^{-+}_{0}\omega^{-A}_{a})e^{B}_{b}\delta_{AB}
-F^{-+}_{23}n_{0}+F^{-A}_{23}e^{B}_{0}\delta_{AB}-\epsilon^{ab}\omega^{-A}_{a}\lambda^{B}_{b}\delta_{AB}
\nonumber \\
\approx&\epsilon^{ab}\lambda^{-A}_{a}e^{B}_{b}\delta_{AB}-\epsilon^{ab}(\omega^{-A}_{0,a}-\omega^{-+}_{a}\omega^{-A}_{0}
-\omega^{-C}_{0}\omega^{DA}_{a}\delta_{CD}+\omega^{-C}_{a}\omega^{DA}_{0}\delta_{CD}+\omega^{-+}_{0}\omega^{-A}_{a})e^{B}_{b}\delta_{AB}
-F^{-+}_{23}n_{0}+F^{-A}_{23}e^{B}_{0}\delta_{AB}
\nonumber \\
\approx&\epsilon^{ab}F^{-A}_{0a}e^{B}_{b}\delta_{AB}-F^{-+}_{23}n_{0}+F^{-A}_{23}e^{B}_{0}\delta_{AB}
=(\eta_{IJ}F^{-I}\wedge e^{J})_{023}=0,\label{4.11}
\end{alignat}
Here, \eqref{2.31}, \eqref{4.173}, \eqref{2.33} and \eqref{3.46} have been used.
The integrability condition \eqref{4.11} is a Ricci identity.

\section{Integrability of $l_{0}$}

Similarly, the integrability conditions for $l_0$ require
\begin{alignat}{1}
l_{0,1a}-l_{0,a1}=&0,\label{4.12}\\
\epsilon^{ab}l_{0,ab}=&0.\label{4.14}
\end{alignat}
The left-hand side of \eqref{4.12} is
\begin{alignat}{1}
l_{0,1a}\approx&\omega^{-+}_{0,a}-\omega^{+A}_{1,a}e^{B}_0\delta_{AB}-\omega^{+A}_{1}e^{B}_{0,a}\delta_{AB}
-\omega^{-+}_{1,a}l_{0}-\omega^{-+}_{1}l_{0,a}
\nonumber \\
\approx&\omega^{-+}_{0,a}-\omega^{+A}_{1,a}e^{B}_0\delta_{AB}-\omega^{+A}_{1}e^{B}_{0,a}\delta_{AB}-\omega^{-+}_{1,a}l_{0}
+\omega^{-+}_{1}(\omega^{+A}_{a}e^{B}_0\delta_{AB}-\omega^{+A}_{0}e^{B}_a\delta_{AB}+\omega^{-+}_{a}l_{0})
\nonumber \\
\approx&\omega^{-+}_{0,a}-(\omega^{+A}_{1,a}-\omega^{-+}_{1}\omega^{+A}_{a})e^{B}_0\delta_{AB}-\omega^{+A}_{1}e^{B}_{0,a}\delta_{AB}
-(\omega^{-+}_{1,a}-\omega^{-+}_{1}\omega^{-+}_{a})l_{0}-\omega^{-+}_{1}\omega^{+A}_{0}e^{B}_a\delta_{AB}
\nonumber \\
\approx&\omega^{-+}_{0,a}-(\omega^{+A}_{1,a}-\omega^{-+}_{1}\omega^{+A}_{a})e^{B}_0\delta_{AB}
+(-\lambda^{A}_{a}+\omega^{-A}_{a}l_0+\omega^{+A}_{a}n_{0}+\omega^{AC}_{a}e^{D}_0\delta_{CD}
-\omega^{AC}_{0}e^D_a\delta_{CD})\omega^{+B}_{1}\delta_{AB}
\nonumber \\&
-(\omega^{-+}_{1,a}-\omega^{-+}_{1}\omega^{-+}_{a})l_{0}-\omega^{-+}_{1}\omega^{+A}_{0}e^{B}_a\delta_{AB}
\nonumber \\
\approx&\omega^{-+}_{0,a}-(\omega^{+A}_{1,a}-\omega^{-+}_{1}\omega^{+A}_{a}-\omega^{+C}_{1}\omega^{DA}_{a}\delta_{CD})e^{B}_0\delta_{AB}
-\lambda^{A}_{a}\omega^{+B}_{1}\delta_{AB}+n_{0}\omega^{+A}_{a}\omega^{+B}_{1}\delta_{AB}
\nonumber \\&
-(\omega^{-+}_{1,a}-\omega^{-+}_{1}\omega^{-+}_{a}-\omega^{-A}_{a}\omega^{+B}_{1}\delta_{AB})l_{0}
-(\omega^{+C}_{1}\omega^{DB}_{0}\delta_{CD}+\omega^{-+}_{1}\omega^{+B}_{0})e^A_a\delta_{AB},
\end{alignat}
and the right-hand side of \eqref{4.12} is
\begin{alignat}{1}
l_{0,a1}\approx&\omega^{+A}_{0,1}e^{B}_a\delta_{AB}+\omega^{+B}_{0}e^{A}_{a,1}\delta_{AB}-\omega^{+A}_{a,1}e^{B}_0\delta_{AB}
-\omega^{+A}_{a}e^{B}_{0,1}\delta_{AB}-\omega^{-+}_{a,1}l_{0}-\omega^{-+}_{a}l_{0,1}
\nonumber \\
\approx&\omega^{+A}_{0,1}e^{B}_a\delta_{AB}+(\omega^{-A}_{a}-\omega^{AC}_{1}e^{D}_{a}\delta_{CD})\omega^{+B}_{0}\delta_{AB}
-\omega^{+A}_{a,1}e^{B}_0\delta_{AB}-\omega^{+A}_{a}(\omega^{-B}_{0}-\omega^{+B}_{1}n_{0}-\omega^{BC}_{1}e^{D}_{0}\delta_{CD})\delta_{AB}
\nonumber \\&
-\omega^{-+}_{a,1}l_{0}-\omega^{-+}_{a}(\omega^{-+}_{0}-\omega^{+A}_{1}e^{B}_0\delta_{AB}-\omega^{-+}_{1}l_{0})
\nonumber \\
\approx&\omega^{+A}_{0,1}e^{B}_a\delta_{AB}+(\omega^{-A}_{a}-\omega^{AC}_{1}e^{D}_{a}\delta_{CD})\omega^{+B}_{0}\delta_{AB}
-(\omega^{+A}_{a,1}-\omega^{+C}_{a}\omega^{DA}_{1}\delta_{CD}-\omega^{-+}_{a}\omega^{+A}_{1})e^{B}_0\delta_{AB}
\nonumber \\&
-(\omega^{-+}_{0}\omega^{-+}_{a}+\omega^{-A}_{0}\omega^{+B}_{a}\delta_{AB})+\omega^{+A}_{a}\omega^{+B}_{1}n_{0}\delta_{AB}
-(\omega^{-+}_{a,1}-\omega^{-+}_{a}\omega^{-+}_{1})l_{0}.
\end{alignat}
\eqref{4.12} requires
\begin{alignat}{1}
&\omega^{-+}_{0,a}-(\omega^{+A}_{1,a}-\omega^{-+}_{1}\omega^{+A}_{a})e^{B}_0\delta_{AB}-\omega^{+A}_{1}e^{B}_{0,a}\delta_{AB}
-(\omega^{-+}_{1,a}-\omega^{-+}_{1}\omega^{-+}_{a})l_{0}-\omega^{-+}_{1}\omega^{+A}_{0}e^{B}_a\delta_{AB}
\nonumber \\
\approx&\omega^{+A}_{0,1}e^{B}_a\delta_{AB}+(\omega^{-A}_{a}-\omega^{AC}_{1}e^{D}_{a}\delta_{CD})\omega^{+B}_{0}\delta_{AB}
-(\omega^{+A}_{a,1}-\omega^{+C}_{a}\omega^{DA}_{1}\delta_{CD}-\omega^{-+}_{a}\omega^{+A}_{1})e^{B}_0\delta_{AB}
\nonumber \\&
-(\omega^{-+}_{0}\omega^{-+}_{a}+\omega^{-A}_{0}\omega^{+B}_{a}\delta_{AB})+\omega^{+A}_{a}\omega^{+B}_{1}n_{0}\delta_{AB}
-(\omega^{-+}_{a,1}-\omega^{-+}_{a}\omega^{-+}_{1})l_{0},
\end{alignat}
which is equivalent to
\begin{alignat}{1}
&\omega^{-+}_{0,a}+(\omega^{+A}_{a,1}-\omega^{+C}_{a}\omega^{DA}_{1}\delta_{CD}-\omega^{-+}_{a}\omega^{+A}_{1}
-\omega^{+A}_{1,a}+\omega^{-+}_{1}\omega^{+A}_{a})e^{B}_0\delta_{AB}-\omega^{+A}_{1}e^{B}_{0,a}\delta_{AB}
+\omega^{-+}_{a,1}l_{0}-\omega^{-+}_{1,a}l_{0}
\nonumber \\&
-\omega^{-+}_{1}\omega^{+A}_{0}e^{B}_a\delta_{AB}-\omega^{+A}_{0,1}e^{B}_a\delta_{AB}
-(\omega^{-A}_{a}-\omega^{AC}_{1}e^{D}_{a}\delta_{CD})\omega^{+B}_{0}\delta_{AB}+\omega^{+A}_{a}\omega^{-B}_{0}\delta_{AB}
-\omega^{+A}_{a}\omega^{+B}_{1}n_{0}\delta_{AB}+\omega^{-+}_{a}\omega^{-+}_{0}
\nonumber \\
\approx&-(\omega^{+A}_{0,1}+\omega^{-+}_{1}\omega^{+A}_{0}-\omega^{+C}_{0}\omega^{DA}_{1}\delta_{CD})e^{B}_a\delta_{AB}
+(\omega^{-+}_{0,a}+\omega^{-+}_{a}\omega^{-+}_{0})-\omega^{+A}_{1}e^{B}_{0,a}\delta_{AB}
-\omega^{+A}_{a}\omega^{+B}_{1}n_{0}\delta_{AB}
\nonumber \\&
-\omega^{-A}_{a}\omega^{+B}_{0}\delta_{AB}+\omega^{+A}_{a}\omega^{-B}_{0}\delta_{AB}+(\omega^{-+}_{a,1}-\omega^{-+}_{1,a})l_{0}
+(\omega^{+A}_{a,1}-\omega^{+A}_{1,a}-\omega^{+C}_{a}\omega^{DA}_{1}\delta_{CD}-\omega^{-+}_{a}\omega^{+A}_{1}
+\omega^{-+}_{1}\omega^{+A}_{a})e^{B}_0\delta_{AB}
\nonumber \\
\approx&-(\omega^{+A}_{0,1}+\omega^{-+}_{1}\omega^{+A}_{0}-\omega^{+C}_{0}\omega^{DA}_{1}\delta_{CD})e^{B}_a\delta_{AB}
+(\omega^{-+}_{0,a}+\omega^{-+}_{a}\omega^{-+}_{0})-\omega^{+A}_{1}e^{B}_{0,a}\delta_{AB}
-\omega^{+A}_{a}\omega^{+B}_{1}n_{0}\delta_{AB}
\nonumber \\&
-\omega^{-A}_{a}\omega^{+B}_{0}\delta_{AB}+\omega^{+A}_{a}\omega^{-B}_{0}\delta_{AB}+F^{-+}_{1a}l_{0}
-\omega^{-A}_{a}\omega^{+B}_{1}\delta_{AB}l_{0}+F^{+A}_{1a}e^{B}_0\delta_{AB}-\omega^{+C}_{1}\omega^{DA}_{a}\delta_{CD}e^{B}_{0}\delta_{AB}
\nonumber \\
\approx&-(e^{A}_{0,a}+\omega^{+A}_{a}n_{0}+\omega^{-A}_{a}l_{0}+\omega^{AC}_{a}e^{D}_{0}\delta_{CD})\omega^{+B}_{1}\delta_{AB}
-(\omega^{+A}_{0,1}+\omega^{-+}_{1}\omega^{+A}_{0}-\omega^{+C}_{0}\omega^{DA}_{1}\delta_{CD})e^{B}_a\delta_{AB}
\nonumber \\&
+(\omega^{-+}_{0,a}+\omega^{-+}_{a}\omega^{-+}_{0}-\omega^{-A}_{a}\omega^{+B}_{0}\delta_{AB}+\omega^{+A}_{a}\omega^{-B}_{0}\delta_{AB})
+F^{-+}_{1a}l_{0}+F^{+A}_{1a}e^{B}_0\delta_{AB}
\nonumber \\
\approx&-(e^{A}_{0,a}+\omega^{+A}_{a}n_{0}+\omega^{-A}_{a}l_{0}+\omega^{AC}_{a}e^{D}_{0}\delta_{CD}
-\omega^{AC}_{0}e^{D}_{a}\delta_{CD})\omega^{+B}_{1}\delta_{AB}
\nonumber \\&
-(\omega^{+A}_{0,1}+\omega^{-+}_{1}\omega^{+A}_{0}-\omega^{+C}_{0}\omega^{DA}_{1}\delta_{CD}
+\omega^{+C}_{1}\omega^{DA}_{0}\delta_{CD}-\omega^{-+}_{0}\omega^{+A}_{1})e^{B}_a\delta_{AB}
\nonumber \\&
+(\omega^{-+}_{0,a}-\omega^{-A}_{a}\omega^{+B}_{0}\delta_{AB}+\omega^{+A}_{a}\omega^{-B}_{0}\delta_{AB})
+F^{-+}_{1a}l_{0}+F^{+A}_{1a}e^{B}_0\delta_{AB}
\nonumber \\
=&-\omega^{+A}_{1}\lambda^{B}_{a}\delta_{AB}-(\omega^{+A}_{0,1}+\omega^{-+}_{1}\omega^{+A}_{0}-\omega^{+C}_{0}\omega^{DA}_{1}\delta_{CD}
+\omega^{+C}_{1}\omega^{DA}_{0}\delta_{CD}-\omega^{-+}_{0}\omega^{+A}_{1})e^{B}_a\delta_{AB}
\nonumber \\&
+(\omega^{-+}_{0,a}-\omega^{-A}_{a}\omega^{+B}_{0}\delta_{AB}+\omega^{+A}_{a}\omega^{-B}_{0}\delta_{AB})
+F^{-+}_{1a}l_{0}+F^{+A}_{1a}e^{B}_0\delta_{AB}
\nonumber \\
\approx&\lambda^{+A}_{1}e^{B}_{a}\delta_{AB}-(\omega^{+A}_{0,1}+\omega^{-+}_{1}\omega^{+A}_{0}
-\omega^{+C}_{0}\omega^{DA}_{1}\delta_{CD}+\omega^{+C}_{1}\omega^{DA}_{0}\delta_{CD}-\omega^{-+}_{0}\omega^{+A}_{1})e^{B}_a\delta_{AB}
\nonumber \\&
-\lambda^{-+}_{a}+(\omega^{-+}_{0,a}-\omega^{-A}_{a}\omega^{+B}_{0}\delta_{AB}+\omega^{+A}_{a}\omega^{-B}_{0}\delta_{AB})
+F^{-+}_{1a}l_{0}+F^{+A}_{1a}e^{B}_0\delta_{AB}
\nonumber \\
\approx&F^{+A}_{01}e^{B}_a\delta_{AB}-F^{-+}_{0a}+F^{-+}_{1a}l_{0}+F^{+A}_{1a}e^{B}_0\delta_{AB}
=(\eta_{IJ}F^{+I}\wedge e^{J})_{01a}=0.\label{4.18}
\end{alignat}
Here, \eqref{4.173}, \eqref{2.48}, \eqref{3.69} and \eqref{3.75} have been used. The integrability conditions \eqref{4.18}
are Ricci identities.

From \eqref{2.26}, one gets
\begin{alignat}{1}
l_{0,ab}\approx&\omega^{+A}_{0,b}e^{B}_a\delta_{AB}+\omega^{+A}_{0}e^{B}_{a,b}\delta_{AB}-\omega^{+A}_{a,b}e^{B}_0\delta_{AB}
-\omega^{+A}_{a}e^{B}_{0,b}\delta_{AB}-\omega^{-+}_{a,b}l_{0}-\omega^{-+}_{a}l_{0,b}.
\end{alignat}
\eqref{4.14} requires
\begin{alignat}{1}
\epsilon^{ab}l_{0,ab}\approx&\epsilon^{ab}(\omega^{+A}_{0,b}e^{B}_a\delta_{AB}+\omega^{+A}_{0}e^{B}_{a,b}\delta_{AB}
-\omega^{+A}_{a,b}e^{B}_0\delta_{AB}-\omega^{+A}_{a}e^{B}_{0,b}\delta_{AB}-\omega^{-+}_{a,b}l_{0}-\omega^{-+}_{a}l_{0,b})
\nonumber \\
\approx&\epsilon^{ab}[\omega^{+A}_{0,b}e^{B}_a\delta_{AB}+\omega^{+A}_{0}e^{B}_{a,b}\delta_{AB}-\omega^{+A}_{a,b}e^{B}_0\delta_{AB}
-\omega^{+A}_{a}e^{B}_{0,b}\delta_{AB}-\omega^{-+}_{a,b}l_{0}
-\omega^{-+}_{a}(\omega^{+A}_{0}e^{B}_b\delta_{AB}-\omega^{+A}_{b}e^{B}_0\delta_{AB}-\omega^{-+}_{b}l_{0})]
\nonumber \\
\approx&-\epsilon^{ab}\omega^{-+}_{a,b}l_{0}+\epsilon^{ab}(-\omega^{+A}_{a,b}+\omega^{-+}_{a}\omega^{+A}_{b})e^{B}_0\delta_{AB}
-\epsilon^{ab}\omega^{+A}_{a}e^{B}_{0,b}\delta_{AB}-\epsilon^{ab}\omega^{+A}_{0,a}e^{B}_b\delta_{AB}
-\epsilon^{ab}\omega^{+A}_{0}(e^{B}_{b,a}+\omega^{-+}_{a}e^{B}_b)\delta_{AB}
\nonumber \\
\approx&-\epsilon^{ab}\omega^{-+}_{a,b}l_{0}+\epsilon^{ab}(-\omega^{+A}_{a,b}+\omega^{-+}_{a}\omega^{+A}_{b})e^{B}_0\delta_{AB}
-\epsilon^{ab}\omega^{+A}_{a}e^{B}_{0,b}\delta_{AB}-\epsilon^{ab}\omega^{+A}_{0,a}e^{B}_b\delta_{AB}
+\epsilon^{ab}\omega^{BC}_{a}e^{D}_{b}\delta_{CD}\omega^{+A}_{0}\delta_{AB}
\nonumber \\&
-\epsilon^{ab}\omega^{-+}_{a}e^{B}_b\omega^{+A}_{0}\delta_{AB}
\nonumber \\
\approx&-\epsilon^{ab}\omega^{-+}_{a,b}l_{0}+\epsilon^{ab}(-\omega^{+A}_{a,b}+\omega^{-+}_{a}\omega^{+A}_{b})e^{B}_0\delta_{AB}
-\epsilon^{ab}\omega^{+A}_{a}e^{B}_{0,b}\delta_{AB}
+\epsilon^{ab}(-\omega^{+A}_{0,a}+\omega^{+C}_{0}\omega^{DA}_{a}\delta_{CD}-\omega^{-+}_{a}\omega^{+A}_{0})e^{B}_b\delta_{AB}
\nonumber \\
\approx&F^{-+}_{23}l_{0}+F^{+A}_{23}e^{B}_0\delta_{AB}-\epsilon^{ab}\omega^{+A}_{a}\lambda^{B}_{b}\delta_{AB}
+\epsilon^{ab}(-\omega^{+A}_{0,a}+\omega^{+C}_{0}\omega^{DA}_{a}\delta_{CD}-\omega^{-+}_{a}\omega^{+A}_{0})e^{B}_b\delta_{AB}\approx0,
\nonumber \\
\approx&F^{-+}_{23}l_{0}+F^{+A}_{23}e^{B}_0\delta_{AB}+\epsilon^{ab}(\lambda^{+A}_{a}-\omega^{+A}_{0,a}
+\omega^{+C}_{0}\omega^{DA}_{a}\delta_{CD}-\omega^{-+}_{a}\omega^{+A}_{0})e^{B}_b\delta_{AB}
\nonumber \\
\approx&F^{-+}_{23}l_{0}+F^{+A}_{23}e^{B}_0\delta_{AB}+\epsilon^{ab}F^{+A}_{0a}e^{B}_b\delta_{AB}
=(\eta_{IJ}F^{+I}\wedge e^{J})_{023}=0,\label{4.16}
\end{alignat}
where \eqref{3.75} have been used. The integrability condition \eqref{4.16} is a Ricci identity.

\section{Integrability of $e^{A}_{0}$}

Finally, the integrability conditions for $e^A_0$ require
\begin{alignat}{1}
e^{A}_{0,1a}-e^{A}_{0,a1}=&0,\label{4.21}\\
\epsilon^{ab}e^{A}_{0,ab}=&0.\label{4.23}
\end{alignat}
From \eqref{2.27} and \eqref{4.173}, one has
\begin{alignat}{1}
e^{A}_{0,1a}\approx&\omega^{-A}_{0,a}-\omega^{+A}_{1,a}n_{0}-\omega^{+A}_{1}n_{0,a}
-\omega^{AB}_{1,a}e^{C}_{0}\delta_{BC}-\omega^{AB}_{1}e^{C}_{0,a}\delta_{BC}
\nonumber \\
\approx&\omega^{-A}_{0,a}-\omega^{+A}_{1,a}n_{0}-\omega^{AB}_{1,a}e^{C}_{0}\delta_{BC}-\omega^{AB}_{1}\delta_{BC}
(\lambda^{C}_{a}-\omega^{-C}_{a}l_0-\omega^{+C}_{a}n_{0}-\omega^{CD}_{a}e^{E}_0\delta_{DE}+\omega^{CD}_{0}e^E_a\delta_{DE})
\nonumber \\&
-\omega^{+A}_{1}(\omega^{-+}_{a}n_{0}+\omega^{-B}_{0}e^{C}_{a}\delta_{BC}-\omega^{-B}_{a}e^{C}_{0}\delta_{BC}),
\end{alignat}
\begin{alignat}{1}
e^{A}_{0,a1}\approx&\lambda^{A}_{a,1}-\omega^{-A}_{a,1}l_0-\omega^{-A}_{a}l_{0,1}-\omega^{+A}_{a,1}n_{0}-\omega^{+A}_{a}n_{0,1}
-\omega^{AB}_{a,1}e^{C}_0\delta_{BC}-\omega^{AB}_{a}e^{C}_{0,1}\delta_{BC}
+\omega^{AB}_{0,1}e^C_a\delta_{BC}+\omega^{AB}_{0}e^C_{a,1}\delta_{BC}
\nonumber \\
\approx&\lambda^{A}_{a,1}-\omega^{-A}_{a,1}l_0-\omega^{-A}_{a}(\omega^{-+}_{0}-\omega^{+A}_{1}e^{B}_0\delta_{AB}-\omega^{-+}_{1}l_{0})
-\omega^{+A}_{a,1}n_{0}-\omega^{+A}_{a}\omega^{-+}_{1}n_{0}-\omega^{AB}_{a,1}e^{C}_0\delta_{BC}
\nonumber \\&
-\omega^{AB}_{a}(\omega^{-C}_{0}-\omega^{+C}_{1}n_{0}-\omega^{CD}_{1}e^{E}_{0}\delta_{DE})\delta_{BC}
+\omega^{AB}_{0,1}e^C_a\delta_{BC}+\omega^{AB}_{0}(\omega^{-C}_{a}-\omega^{CD}_{1}e^{E}_{a}\delta_{DE})\delta_{BC}.
\end{alignat}
\eqref{4.21} requires
\begin{alignat}{1}
&\omega^{-A}_{0,a}-\omega^{+A}_{1,a}n_{0}-\omega^{AB}_{1,a}e^{C}_{0}\delta_{BC}-\omega^{AB}_{1}\delta_{BC}
(\lambda^{C}_{a}-\omega^{-C}_{a}l_0-\omega^{+C}_{a}n_{0}-\omega^{CD}_{a}e^{E}_0\delta_{DE}+\omega^{CD}_{0}e^E_a\delta_{DE})
\nonumber \\&
-\omega^{+A}_{1}(\omega^{-+}_{a}n_{0}+\omega^{-B}_{0}e^{C}_{a}\delta_{BC}-\omega^{-B}_{a}e^{C}_{0}\delta_{BC})
-\lambda^{A}_{a,1}+\omega^{-A}_{a,1}l_0+\omega^{-A}_{a}(\omega^{-+}_{0}-\omega^{+B}_{1}e^{C}_0\delta_{BC}-\omega^{-+}_{1}l_{0})
\nonumber \\
&+\omega^{+A}_{a,1}n_{0}+\omega^{+A}_{a}\omega^{-+}_{1}n_{0}+\omega^{AB}_{a,1}e^{C}_0\delta_{BC}
+\omega^{AB}_{a}(\omega^{-C}_{0}-\omega^{+C}_{1}n_{0}-\omega^{CD}_{1}e^{E}_{0}\delta_{DE})\delta_{BC}
-\omega^{AB}_{0,1}e^C_a\delta_{BC}
\nonumber \\&
-\omega^{AB}_{0}(\omega^{-C}_{a}-\omega^{CD}_{1}e^{E}_{a}\delta_{DE})\delta_{BC}
\nonumber \\
\approx&\omega^{-A}_{0,a}-\omega^{+A}_{1,a}n_{0}-\omega^{AB}_{1,a}e^{C}_{0}\delta_{BC}
-\omega^{AB}_{1}\delta_{BC}\lambda^{C}_{a}+\omega^{AB}_{1}\delta_{BC}\omega^{-C}_{a}l_0+\omega^{AB}_{1}\delta_{BC}\omega^{+C}_{a}n_{0}
+\omega^{AB}_{1}\delta_{BC}\omega^{CD}_{a}e^{E}_0\delta_{DE}
\nonumber \\&
-\omega^{AB}_{1}\delta_{BC}\omega^{CD}_{0}e^E_a\delta_{DE}-\omega^{+A}_{1}\omega^{-+}_{a}n_{0}-\omega^{+A}_{1}\omega^{-B}_{0}e^{C}_{a}\delta_{BC}
+\omega^{+A}_{1}\omega^{-B}_{a}e^{C}_{0}\delta_{BC}-\lambda^{A}_{a,1}+\omega^{-A}_{a,1}l_0+\omega^{-A}_{a}\omega^{-+}_{0}
\nonumber \\&
-\omega^{-A}_{a}\omega^{+B}_{1}e^{C}_0\delta_{BC}-\omega^{-A}_{a}\omega^{-+}_{1}l_{0}+\omega^{+A}_{a,1}n_{0}
+\omega^{+A}_{a}\omega^{-+}_{1}n_{0}+\omega^{AB}_{a,1}e^{C}_0\delta_{BC}+\omega^{AB}_{a}\omega^{-C}_{0}\delta_{BC}
-\omega^{AB}_{a}\omega^{+C}_{1}n_{0}\delta_{BC}
\nonumber \\&
-\omega^{AB}_{a}\omega^{CD}_{1}e^{E}_{0}\delta_{DE}\delta_{BC}-\omega^{AB}_{0,1}e^C_a\delta_{BC}
-\omega^{AB}_{0}\omega^{-C}_{a}\delta_{BC}+\omega^{AB}_{0}\omega^{CD}_{1}e^{E}_{a}\delta_{DE}\delta_{BC}
\nonumber \\
\approx&(\omega^{-A}_{0,a}+\omega^{-A}_{a}\omega^{-+}_{0}+\omega^{AB}_{a}\omega^{-C}_{0}\delta_{BC}-\omega^{AB}_{0}\omega^{-C}_{a}\delta_{BC}
-\lambda^{-A}_{a})+(\omega^{-A}_{a,1}-\omega^{-A}_{a}\omega^{-+}_{1}+\omega^{AB}_{1}\omega^{-C}_{a}\delta_{BC})l_0
\nonumber \\&
+(\omega^{+A}_{a,1}-\omega^{+A}_{1,a}+\omega^{AB}_{1}\omega^{+C}_{a}\delta_{BC}-\omega^{+A}_{1}\omega^{-+}_{a}
+\omega^{+A}_{a}\omega^{-+}_{1}-\omega^{AB}_{a}\omega^{+C}_{1}\delta_{BC})n_{0}
\nonumber \\&
+(\omega^{AB}_{a,1}-\omega^{AB}_{1,a}+\omega^{AD}_{1}\delta_{DE}\omega^{EB}_{a}+\omega^{+A}_{1}\omega^{-B}_{a}
-\omega^{-A}_{a}\omega^{+B}_{1}-\omega^{AD}_{a}\omega^{EB}_{1}\delta_{DE})e^{C}_0\delta_{BC}
\nonumber \\&
+(\lambda^{AB}_{1}-\omega^{AB}_{0,1}-\omega^{AD}_{1}\omega^{EB}_{0}\delta_{DE}-\omega^{+A}_{1}\omega^{-B}_{0}
+\omega^{AD}_{0}\omega^{EB}_{1}\delta_{DE})e^{C}_{a}\delta_{BC}
\nonumber \\
\approx&(\omega^{-A}_{0,a}+\omega^{-+}_{0}\omega^{-A}_{a}-\omega^{-+}_{a}\omega^{-A}_{0}-\omega^{-B}_{0}\omega^{CA}_{a}\delta_{BC}
+\omega^{-B}_{a}\omega^{CA}_{0}\delta_{BC}-\lambda^{-A}_{a})+F^{-A}_{1a}l_0+F^{+A}_{1a}n_{0}+F^{AB}_{1a}e^{C}_0\delta_{BC}
\nonumber \\&
+(\lambda^{AB}_{1}-\omega^{AB}_{0,1}+\omega^{-A}_{0}\omega^{+B}_{1}-\omega^{AD}_{1}\omega^{EB}_{0}\delta_{DE}-\omega^{+A}_{1}\omega^{-B}_{0}
+\omega^{AD}_{0}\omega^{EB}_{1}\delta_{DE})e^{C}_{a}\delta_{BC}
\nonumber \\
\approx&-F^{-A}_{0a}+F^{-A}_{1a}l_0+F^{+A}_{1a}n_{0}+F^{AB}_{1a}e^{C}_0\delta_{BC}+F^{AB}_{01}e^{C}_{a}\delta_{BC}
=(\eta_{IJ}F^{AI}\wedge e^{J})_{01a}=0,\label{4.21-1}
\end{alignat}
where \eqref{3.49} has been used. The integrability conditions \eqref{4.21-1} are Ricci identities.

From \eqref{4.173}, one gets
\begin{alignat}{1}
e^{A}_{0,ab}\approx&\lambda^{A}_{a,b}-\omega^{-A}_{a,b}l_0-\omega^{-A}_{a}l_{0,b}-\omega^{+A}_{a,b}n_{0}-\omega^{+A}_{a}n_{0,b}
-\omega^{AB}_{a,b}e^{C}_0\delta_{BC}-\omega^{AB}_{a}e^{C}_{0,b}\delta_{BC}+\omega^{AB}_{0,b}e^C_a\delta_{BC}+\omega^{AB}_{0}e^C_{a,b}\delta_{BC}.
\end{alignat}
\eqref{4.23} require
\begin{alignat}{1}
\epsilon^{ab}e^{A}_{0,ab}\approx&\epsilon^{ab}\lambda^{A}_{a,b}-\epsilon^{ab}\omega^{-A}_{a,b}l_0-\epsilon^{ab}\omega^{-A}_{a}l_{0,b}
-\epsilon^{ab}\omega^{+A}_{a,b}n_{0}-\epsilon^{ab}\omega^{+A}_{a}n_{0,b}-\epsilon^{ab}\omega^{AB}_{a,b}e^{C}_0\delta_{BC}
-\epsilon^{ab}\omega^{AB}_{a}e^{C}_{0,b}\delta_{BC}
\nonumber \\&
+\epsilon^{ab}\omega^{AB}_{0,b}e^C_a\delta_{BC}+\epsilon^{ab}\omega^{AB}_{0}e^C_{a,b}\delta_{BC}
\nonumber \\
\approx&\lambda^{A}_{a,b}-\omega^{-A}_{a,b}l_0+\omega^{-A}_{a}(\omega^{+B}_{b}e^{C}_0\delta_{BC}-\omega^{+B}_{0}e^{C}_b\delta_{BC}+\omega^{-+}_{b}l_{0})
-\omega^{+A}_{a,b}n_{0}-\omega^{+A}_{a}(\omega^{-+}_{b}n_{0}+\omega^{-B}_{0}e^{C}_{b}\delta_{BC}-\omega^{-B}_{b}e^{C}_{0}\delta_{BC})
\nonumber \\&
-\omega^{AB}_{a,b}e^{C}_0\delta_{BC}-\omega^{AB}_{a}(\lambda^{C}_{b}-\omega^{-C}_{b}l_0-\omega^{+C}_{b}n_{0}
-\omega^{CD}_{b}e^{E}_0\delta_{DE}-\omega^{CD}_{0}e^E_b\delta_{DE})\delta_{BC}
+\omega^{AB}_{0,b}e^C_a\delta_{BC}+\omega^{AB}_{0}e^C_{a,b}\delta_{BC}
\nonumber \\
\approx&\epsilon^{ab}\lambda^{A}_{a,b}-\epsilon^{ab}\omega^{-A}_{a,b}l_0
+\epsilon^{ab}\omega^{-A}_{a}\omega^{+B}_{b}e^{C}_0\delta_{BC}-\epsilon^{ab}\omega^{-A}_{a}\omega^{+B}_{0}e^{C}_b\delta_{BC}
+\epsilon^{ab}\omega^{-A}_{a}\omega^{-+}_{b}l_{0}-\epsilon^{ab}\omega^{+A}_{a,b}n_{0}
\nonumber \\&
-\epsilon^{ab}\omega^{+A}_{a}\omega^{-+}_{b}n_{0}-\epsilon^{ab}\omega^{+A}_{a}\omega^{-B}_{0}e^{C}_{b}\delta_{BC}
+\epsilon^{ab}\omega^{+A}_{a}\omega^{-B}_{b}e^{C}_{0}\delta_{BC}
-\epsilon^{ab}\omega^{AB}_{a,b}e^{C}_0\delta_{BC}-\epsilon^{ab}\omega^{AB}_{a}\lambda^{C}_{b}\delta_{BC}
\nonumber \\&
+\epsilon^{ab}\omega^{AB}_{a}\omega^{-C}_{b}l_0\delta_{BC}+\epsilon^{ab}\omega^{AB}_{a}\omega^{+C}_{b}n_{0}\delta_{BC}
+\epsilon^{ab}\omega^{AB}_{a}\omega^{CD}_{b}e^{E}_0\delta_{DE}\delta_{BC}
+\epsilon^{ab}\omega^{AB}_{a}\omega^{CD}_{0}e^E_b\delta_{DE}\delta_{BC}
\nonumber \\&
+\epsilon^{ab}\omega^{AB}_{0,b}e^C_a\delta_{BC}+\epsilon^{ab}\omega^{AB}_{0}\omega^{CD}_{a}e^{E}_{b}\delta_{DE}\delta_{BC}
\nonumber \\
\approx&(-\epsilon^{ab}\omega^{+A}_{a,b}-\epsilon^{ab}\omega^{+C}_{a}\omega^{AB}_{b}\delta_{BC}+\epsilon^{ab}\omega^{-+}_{a}\omega^{+A}_{b})n_{0}
+(-\epsilon^{ab}\omega^{-A}_{a,b}-\epsilon^{ab}\omega^{-+}_{a}\omega^{-A}_{b}+\epsilon^{ab}\omega^{-B}_{a}\omega^{CA}_{b}\delta_{BC})l_0
\nonumber \\&
+(\epsilon^{ab}\omega^{-A}_{a}\omega^{+B}_{b}+\epsilon^{ab}\omega^{+A}_{a}\omega^{-B}_{b}
-\epsilon^{ab}\omega^{AB}_{a,b}+\epsilon^{ab}\omega^{AD}_{a}\omega^{EB}_{b}\delta_{DE})e^{C}_0\delta_{BC}
\nonumber \\&
+(\epsilon^{ab}\lambda^{AB}_{a}-\epsilon^{ab}\omega^{-A}_{a}\omega^{+B}_{0}
-\epsilon^{ab}\omega^{+A}_{a}\omega^{-B}_{0}+\epsilon^{ab}\omega^{AD}_{a}\omega^{EB}_{0}\delta_{DE}
-\epsilon^{ab}\omega^{AB}_{0,a}+\epsilon^{ab}\omega^{AD}_{0}\omega^{EB}_{a}\delta_{DE})e^{C}_{b}\delta_{BC}
\nonumber \\
\approx&\epsilon^{ab}(\lambda^{AB}_{a}-\omega^{AB}_{0,a}-\omega^{-A}_{a}\omega^{+B}_{0}-\omega^{+A}_{a}\omega^{-B}_{0}
+\omega^{AD}_{a}\omega^{EB}_{0}\delta_{DE}+\omega^{AD}_{0}\omega^{EB}_{a}\delta_{DE})e^{C}_{b}\delta_{BC}
+F^{+A}_{23}n_{0}+F^{-A}_{23}l_0+F^{AB}_{23}e^{C}_0\delta_{BC}
\nonumber \\
\approx&\epsilon^{ab}F^{AB}_{0a}e^{C}_{b}\delta_{BC}+F^{+A}_{23}n_{0}+F^{-A}_{23}l_0+F^{AB}_{23}e^{C}_0\delta_{BC}
=(\eta_{IJ}F^{AI}\wedge e^{J})_{023}=0,\label{4.23-1}
\end{alignat}
where \eqref{2.31}, \eqref{2.26}, \eqref{4.173}, \eqref{2.33} and \eqref{3.50} have been used.
The integrability condition \eqref{4.23-1} is a Ricci identity.

\section{Poisson Brackets Among Constraints}

All non-zero Poisson brackets among constraints are listed as follows:
\begin{alignat}{1}
&\{\pi^{a}_{A}(x),(\pi^{1}_{-+}-2\epsilon_{BC}\epsilon^{bc}e^{B}_{b}e^{C}_{c})(y)\}
=4\epsilon_{AB}\epsilon^{ab}e^{B}_{b}(y)\delta(x-y),\\
&\{\pi^{a}_{A}(x),(\pi^{b}_{-B}-4\epsilon_{BC}\epsilon^{bc}e^{C}_{c})(y)\}=4\epsilon_{AB}\epsilon^{ab}\delta(x-y),\\
&\{\pi^{a}_{A}(x),(\epsilon^{bc}\omega^{-B}_{b}e^C_c\delta_{BC})(y)\}=\epsilon^{ab}\omega^{-A}_{b}(y)\delta(x-y),\\
&\{\pi^{a}_{A}(x),(\omega^{-+}_{b}-\omega^{+B}_{1}e^{C}_{b}\delta_{BC})(y)\}=\omega^{+A}_{1}(y)\delta(x-y)\delta^{a}_{b},\\
&\{\pi^{a}_{A}(x),(\epsilon^{bc}\omega^{+B}_{b}e^{C}_c\delta_{BC})(y)\}=\epsilon^{ab}\omega^{+A}_{b}(y)\delta(x-y),\\
&\{\pi^{a}_{A}(x),(e^{B}_{b,1}-\omega^{-B}_{b}+\omega^{BC}_{1}e^{D}_{b}\delta_{CD})(y)\}
=\omega^{AB}_{1}(y)\delta(x-y)\delta^{a}_{b}-\delta(x-y)_{,y^{1}}\delta^{B}_{A}\delta^{a}_{b},\\
&\{\pi^{a}_{A}(x),\epsilon^{bc}(e^{B}_{b,c}-\omega^{BC}_{b}e^{D}_{c}\delta_{CD})(y)\}
=\epsilon^{ab}\omega^{+A}_{b}(y)\delta(x-y)-\epsilon^{ab}\delta(x-y)_{,y^{b}}\delta^{B}_{A},\\
&\{\pi^{a}_{A}(x),(\epsilon_{BC}\epsilon^{bc}F^{-B}_{1b}e^{C}_{c})(y)\}=\epsilon_{AB}\epsilon^{ab}F^{-B}_{1b}(y)\delta(x-y),\\
&\{\pi^{a}_{A}(x),(\epsilon_{BC}\epsilon^{bc}F^{+B}_{1b}e^{C}_{c}+F^{23}_{23})(y)\}=\epsilon_{AB}\epsilon^{ab}F^{+B}_{1b}(y)\delta(x-y),\\
&\{\pi^{a}_{A}(x),(F^{-B}_{23}+\epsilon^{bc}e^{B}_bF^{-+}_{1c})(y)\}=\epsilon^{ab}F^{-+}_{1b}(y)\delta(x-y)\delta^{B}_{A},\\
&\{\pi^{a}_{A}(x),(\epsilon^{bc}F^{+2}_{1b}e^{3}_{c}+\epsilon^{bc}F^{+3}_{1b}e^{2}_{c})(y)\}
=-\epsilon^{ab}F^{+2}_{1b}(y)\delta(x-y)\delta^{3}_{A}-\epsilon^{ab}F^{+3}_{1b}(y)\delta(x-y)\delta^{2}_{A},\\
&\{\pi^{a}_{A}(x),(\epsilon^{bc}F^{+2}_{1b}e^{2}_{c}-\epsilon^{bc}F^{+3}_{1b}e^{3}_{c})(y)\}
=-\epsilon^{ab}F^{+2}_{1b}(y)\delta(x-y)\delta^{2}_{A}+\epsilon^{ab}F^{+3}_{1b}(y)\delta(x-y)\delta^{3}_{A},\\
&\{(\pi^{1}_{-+}-2\epsilon_{AB}\epsilon^{ab}e^{A}_{a}e^{B}_{b})(x),(\epsilon_{CD}\epsilon^{cd}F^{-C}_{1c}e^{D}_d)(y)\}
=\epsilon_{AB}\epsilon^{ab}\omega^{-A}_{a}(y)e^{B}_{b}(y)\delta(x-y),\\
&\{(\pi^{1}_{-+}-2\epsilon_{AB}\epsilon^{ab}e^{A}_{a}e^{B}_{b})(x),(\epsilon_{CD}\epsilon^{cd}F^{+C}_{1c}e^{D}_{d}+F^{23}_{23})(y)\}
=\epsilon_{AB}\epsilon^{ab}\omega^{+A}_{a}(y)e^{B}_{b}(y)\delta(x-y),\\
&\{(\pi^{1}_{-+}-2\epsilon_{AB}\epsilon^{ab}e^{A}_{a}e^{B}_{b})(x),(F^{-C}_{23}+\epsilon^{cd}e^{C}_cF^{-+}_{1d})(y)\}
=\epsilon^{ab}e^{C}_{a}(y)\delta(x-y)_{,y^{b}},\\
&\{(\pi^{1}_{-+}-2\epsilon_{AB}\epsilon^{ab}e^{A}_{a}e^{B}_{b})(x),(\epsilon^{cd}F^{+2}_{1c}e^{3}_{d}+\epsilon^{cd}F^{+3}_{1c}e^{2}_{d})(y)\}
=-[\epsilon^{ab}\omega^{+2}_{a}(y)e^{3}_{b}(y)+\epsilon^{ab}\omega^{+3}_{a}(y)e^{2}_{b}(y)]\delta(x-y),\\
&\{(\pi^{1}_{-+}-2\epsilon_{AB}\epsilon^{ab}e^{A}_{a}e^{B}_{b})(x),(\epsilon^{cd}F^{+2}_{1c}e^{2}_{d}-\epsilon^{cd}F^{+3}_{1c}e^{3}_{d})(y)\}
=\epsilon^{ab}\omega^{+3}_{a}(y)e^{3}_{b}(y)\delta(x-y)-\epsilon^{ab}\omega^{+2}_{a}(y)e^{2}_{b}(y)\delta(x-y),\\
&\{\pi^{a}_{-+}(x),(\omega^{-+}_{b}-\omega^{+A}_{1}e^{B}_{b}\delta_{AB})(y)\}=-\delta(x-y)\delta^{a}_{b},\\
&\{\pi^{a}_{-+}(x),(\epsilon_{AB}\epsilon^{bc}F^{+A}_{1b}e^{B}_{c}+F^{23}_{23})(y)\}
=\epsilon_{AB}\epsilon^{ab}\omega^{+A}_{1}(y)e^{B}_{b}(y)\delta(x-y),\\
&\{\pi^{a}_{-+}(x),(F^{-A}_{23}+\epsilon^{bc}e^{A}_bF^{-+}_{1c})(y)\}=\epsilon^{ab}e^{A}_b(y)\delta(x-y)_{,y^{1}},\\
&\{\pi^{a}_{-+}(x),(\epsilon^{bc}F^{+2}_{1b}e^{3}_{c}+\epsilon^{bc}F^{+3}_{1b}e^{2}_{c})(y)\}
=\epsilon^{ab}\omega^{+2}_{1}(y)e^{3}_{b}(y)\delta(x-y)+\epsilon^{ab}\omega^{+3}_{1}(y)e^{2}_{b}(y)\delta(x-y),\\
&\{\pi^{a}_{-+}(x),(\epsilon^{bc}F^{+2}_{1b}e^{2}_{c}-\epsilon^{bc}F^{+3}_{1b}e^{3}_{c})(y)\}
=\epsilon^{ab}\omega^{+2}_{1}(y)e^{2}_{b}(y)\delta(x-y)-\epsilon^{ab}\omega^{+3}_{1}(y)e^{3}_{b}(y)\delta(x-y),\\
&\{\pi^{1}_{-A}(x),\omega^{-B}_{1}(y)\}=-\delta(x-y)\delta^{B}_{A},\\
&\{(\pi^{a}_{-A}-4\epsilon_{AB}\epsilon^{ab}e^{B}_{b})(x),(\epsilon^{cd}\omega^{-C}_{c}e^D_d\delta_{CD})(y)\}
=-\epsilon^{ab}e^{A}_{b}(y)\delta(x-y),\\
&\{(\pi^{a}_{-A}-4\epsilon_{AB}\epsilon^{ab}e^{B}_{b})(x),(e^{C}_{c,1}-\omega^{-C}_{c}+\omega^{CD}_{1}e^{E}_{c}\delta_{DE})(y)\}
=\delta(x-y)\delta^{C}_{A}\delta^{a}_{c},\\
&\{(\pi^{a}_{-A}-4\epsilon_{AB}\epsilon^{ab}e^{B}_{b})(x),(\epsilon_{CD}\epsilon^{cd}F^{-C}_{1c}e^{D}_d)(y)\}
=\epsilon_{AB}\epsilon^{ab}[\omega^{-+}_{1}(y)e^{B}_b(y)\delta(x-y)-e^{B}_{b}(y)\delta(x-y)_{,y^{1}}],\\
&\{(\pi^{a}_{-A}-4\epsilon_{AB}\epsilon^{ab}e^{B}_{b})(x),(\epsilon_{CD}\epsilon^{cd}F^{+C}_{1c}e^{D}_{d}+F^{23}_{23})(y)\}
=\epsilon^{ab}\omega^{+2}_{b}(y)\delta^{3}_{A}\delta(x-y)-\epsilon^{ab}\omega^{+3}_{b}(y)\delta^{2}_{A}\delta(x-y),\\
&\{(\pi^{a}_{-A}-4\epsilon_{AB}\epsilon^{ab}e^{B}_{b})(x),(F^{-C}_{23}+\epsilon^{cd}e^{C}_cF^{-+}_{1d})(y)\}
=\epsilon^{ab}\omega^{+A}_{1}(y)e^{C}_b(y)\delta(x-y)-\epsilon^{bc}\delta(x-y)_{,y^b}\delta^{C}_{A}\delta^{a}_{c},\\
&\{\pi^{1}_{+A}(x),(\omega^{-+}_{a}-\omega^{+B}_{1}e^{C}_{a}\delta_{BC})(y)\}=e^{A}_{a}(y)\delta(x-y),\\
&\{\pi^{1}_{+A}(x),(\epsilon_{BC}\epsilon^{bc}F^{+B}_{1b}e^{C}_{c}+F^{23}_{23})(y)\}
=[\epsilon^{ab}\omega^{23}_{a}(y)e^{A}_{b}(y)+\epsilon_{AB}\epsilon^{ab}\omega^{-+}_{a}(y)e^{B}_{b}(y)]\delta(x-y)
\nonumber \\&
-\epsilon^{ab}e^{A}_{a}(y)\delta(x-y)_{,y^{b}},\\
&\{\pi^{1}_{+A}(x),(F^{-B}_{23}+\epsilon^{bc}e^{B}_bF^{-+}_{1c})(y)\}
=\epsilon^{ab}\omega^{-A}_{a}(y)e^{B}_{b}(y)\delta(x-y), \\
&\{\pi^{1}_{+A}(x),(\epsilon^{ab}F^{+2}_{1a}e^{3}_{b}+\epsilon^{ab}F^{+3}_{1a}e^{2}_{b})(y)\}
=\epsilon^{ab}\omega^{23}_{a}(y)e^{3}_{b}(y)\delta(x-y)\delta^{3}_{A}+\epsilon^{ab}\omega^{-+}_{a}(y)e^{3}_{b}(y)\delta(x-y)\delta^{2}_{A}
\nonumber \\&
+\epsilon^{ab}e^{2}_{b}(y)\delta(x-y)_{,y^{a}}\delta^{3}_{A}+\epsilon^{ab}\omega^{23}_{a}(y)e^{2}_{b}(y)\delta(x-y)\delta^{2}_{A}
-\epsilon^{ab}\omega^{-+}_{a}(y)e^{2}_{b}(y)\delta(x-y)\delta^{3}_{A}+\epsilon^{ab}e^{3}_{b}(y)\delta(x-y)_{,y^{a}}\delta^{2}_{A},\\
&\{\pi^{1}_{+A}(x),(\epsilon^{ab}F^{+2}_{1a}e^{2}_{b}-\epsilon^{ab}F^{+3}_{1a}e^{3}_{b})(y)\}
=\epsilon^{ab}\omega^{23}_{a}(y)e^{2}_{b}(y)\delta(x-y)\delta^{3}_{A}+\epsilon^{ab}\omega^{-+}_{a}(y)e^{2}_{b}(y)\delta(x-y)\delta^{2}_{A}
\nonumber \\&
-\epsilon^{ab}e^{3}_{b}(y)\delta(x-y)_{,y^{a}}\delta^{3}_{A}-\epsilon^{ab}\omega^{23}_{a}(y)e^{3}_{b}(y)\delta(x-y)\delta^{2}_{A}
+\epsilon^{ab}\omega^{-+}_{a}(y)e^{3}_{b}(y)\delta(x-y)\delta^{3}_{A}+\epsilon^{ab}e^{2}_{b}(y)\delta(x-y)_{,y^{a}}\delta^{2}_{A},\\
&\{\pi^{a}_{+A}(x),\epsilon^{bc}(\omega^{+B}_{b}e^{C}_c\delta_{BC})(y)\}=-\epsilon^{ab}e^{A}_b(y)\delta(x-y),\\
&\{\pi^{a}_{+A}(x),(\epsilon_{BC}\epsilon^{bc}F^{+B}_{1b}e^{C}_{c}+F^{23}_{23})(y)\}
=\epsilon^{ab}\omega^{-2}_{b}(y)\delta^{3}_{A}\delta(x-y)-\epsilon^{ab}\omega^{-3}_{b}(y)\delta^{2}_{A}\delta(x-y)
\nonumber \\&
-\epsilon_{AB}\epsilon^{ab}\omega^{-+}_{1}(y)e^{B}_{b}(y)\delta(x-y)-\epsilon^{ab}\omega^{23}_{1}(y)e^{A}_{b}(y)\delta(x-y)
-\epsilon_{AB}\epsilon^{ab}e^{B}_{b}(y)\delta(x-y)_{,y^{1}},
\end{alignat}
\begin{alignat}{1}
&\{\pi^{a}_{+A}(x),(\epsilon^{bc}F^{+2}_{1b}e^{3}_{c}+\epsilon^{bc}F^{+3}_{1b}e^{2}_{c})(y)\}
=-\epsilon^{ab}\omega^{-+}_{1}(y)e^{3}_{b}(y)\delta(x-y)\delta^{2}_{A}-\epsilon^{ab}\omega^{23}_{1}(y)e^{3}_{b}(y)\delta(x-y)\delta^{3}_{A}
\nonumber \\&
-\epsilon^{ab}e^{2}_{b}(y)\delta(x-y)_{,y^{1}}\delta^{3}_{A}-\epsilon^{ab}\omega^{-+}_{1}(y)e^{2}_{b}(y)\delta(x-y)\delta^{3}_{A}
+\epsilon^{ab}\omega^{23}_{1}(y)e^{2}_{b}(y)\delta(x-y)\delta^{2}_{A}-\epsilon^{ab}e^{3}_{b}(y)\delta(x-y)_{,y^{1}}\delta^{2}_{A},\\
&\{\pi^{a}_{+A}(x),(\epsilon^{bc}F^{+2}_{1b}e^{2}_{c}-\epsilon^{bc}F^{+3}_{1b}e^{3}_{c})(y)\}
=-\epsilon^{ab}\omega^{-+}_{1}(y)e^{2}_{b}(y)\delta(x-y)\delta^{2}_{A}-\epsilon^{ab}\omega^{23}_{1}(y)e^{2}_{b}(y)\delta(x-y)\delta^{3}_{A}
\nonumber \\&
+\epsilon^{ab}e^{3}_{b}(y)\delta(x-y)_{,y^{1}}\delta^{3}_{A}+\epsilon^{ab}\omega^{-+}_{1}(y)e^{3}_{b}(y)\delta(x-y)\delta^{3}_{A}
-\epsilon^{ab}\omega^{23}_{1}(y)e^{3}_{b}(y)\delta(x-y)\delta^{2}_{A}-\epsilon^{ab}e^{2}_{b}(y)\delta(x-y)_{,y^{1}}\delta^{2}_{A},\\
&\{\pi^{1}_{23}(x),(e^{A}_{a,1}-\omega^{-A}_{a}+\omega^{AB}_{1}e^{C}_{a}\delta_{BC})(y)\}
=-\epsilon_{AB}e^{B}_{a}(y)\delta(x-y),\\
&\{\pi^{1}_{23}(x),(\epsilon^{ab}F^{+2}_{1a}e^{3}_{b}+\epsilon^{ab}F^{+3}_{1a}e^{2}_{b})(y)\}
=-\epsilon^{ab}\omega^{+3}_{a}(y)e^{3}_{b}(y)\delta(x-y)+\epsilon^{ab}\omega^{+2}_{a}(y)e^{2}_{b}(y)\delta(x-y),\\
&\{\pi^{1}_{23}(x),(\epsilon^{ab}F^{+2}_{1a}e^{2}_{b}-\epsilon^{ab}F^{+3}_{1a}e^{3}_{b})(y)\}
=-\epsilon^{ab}\omega^{+3}_{a}(y)e^{2}_{b}(y)\delta(x-y)-\epsilon^{ab}\omega^{+2}_{a}(y)e^{3}_{b}(y)\delta(x-y),\\
&\{\pi^{a}_{23}(x),\epsilon^{bc}(e^{A}_{b,c}-\omega^{AB}_{b}e^{C}_{c}\delta_{BC})(y)\}
=\epsilon_{AB}\epsilon^{ab}e^{B}_{b}(y)\delta(x-y),\\
&\{\pi^{a}_{23}(x),(\epsilon_{AB}\epsilon^{bc}F^{+A}_{1b}e^{B}_{c}+F^{23}_{23})(y)\}
=\epsilon^{ab}\omega^{+A}_{1}(y)e^{B}_{b}(y)\delta_{AB}\delta(x-y)+\epsilon^{ab}\delta(x-y)_{,y^{b}},\\
&\{\pi^{a}_{23}(x),(F^{-A}_{23}+\epsilon^{bc}e^{A}_bF^{-+}_{1c})(y)\}
=-\epsilon_{AB}\epsilon^{ab}\omega^{-B}_{b}(y)\delta(x-y),\\
&\{\pi^{a}_{23}(x),(\epsilon^{bc}F^{+2}_{1b}e^{3}_{c}+\epsilon^{bc}F^{+3}_{1b}e^{2}_{c})(y)\}
=\epsilon^{ab}\omega^{+3}_{1}(y)e^{3}_{b}(y)\delta(x-y)-\epsilon^{ab}\omega^{+2}_{1}(y)e^{2}_{b}(y)\delta(x-y),\\
&\{\pi^{a}_{23}(x),(\epsilon^{bc}F^{+2}_{1b}e^{2}_{c}-\epsilon^{bc}F^{+3}_{1b}e^{3}_{c})(y)\}
=\epsilon^{ab}\omega^{+3}_{1}(y)e^{2}_{b}(y)\delta(x-y)+\epsilon^{ab}\omega^{+2}_{1}(y)e^{3}_{b}(y)\delta(x-y).
\end{alignat}


\begin{thebibliography}{10}


\bibitem{ADM}R. Arnowitt, S. Deser and C. Misner, Phys. Rev. $\mathbf{116}$ (1959) 1322.

\bibitem{d'IS}R. A. d'Inverno and J. Smallwood, Phys. Rev. D {\bf 22} (1980) 1233.

\bibitem{Goldberg1}J. Goldberg, Found. Phys. $\mathbf{14}$ (1984) 1211.

\bibitem{Bondi}H. Bondi, M.G.J. van der Burg and A.W.K. Metzner, Proc. R. Soc. Lond. A $\mathbf{269}$ (1962), 21.

\bibitem{Sachs}R.K. Sachs, Proc. R. Soc. Lond. A $\mathbf{270}$ (1962), 103.

\bibitem{ABF}A. Ashtakar, C. Beetle and S. Fairhurst, Class. Quant. Grav. {\bf 16} (1999) L1.

\bibitem{AK}A. Ashtekar and B. Krishnan, Living Rev. Relativ. {\bf 7} (2004) 10.

\bibitem{Krishnan}B. Krishnan (2014) pp 527-555. In: Ashtekar A., Petkov V. (eds) Springer Handbook of Spacetime.
Springer, Berlin, Heidelberg.

\bibitem{Ashtekar}A. Ashtekar, Phys. Rev. Lett. $\mathbf{57}$ (1986) 2244.

\bibitem{Inverno1}R. d'Inverno and J. Vickers, Class. Quant. Grav. $\mathbf{12}$ (1995) 753.

\bibitem{Inverno2}R. d'Inverno, P. Lambert and J. Vickers, Class. Quant. Grav. $\mathbf{23}$ (2006) 3747.

\bibitem{Inverno3}R. d'Inverno, P. Lambert and J. Vickers, Class. Quant. Grav. $\mathbf{23}$ (2006) 4511.

\bibitem{Goldberg2}J. Goldberg, D. Robinson and C. Soteriou, Class. Quant. Grav. $\mathbf{9}$ (1992) 1309.

\bibitem{Goldberg3}J. Goldberg and C. Soteriou, Class. Quant. Grav. $\mathbf{12}$ (1995) 2779.

\bibitem{BTT}N. Bodendorfer, T. Thiemann and A. Thurn, Class. Quant. Grav. $\mathbf{30}$ (2013), 045001, 045002, 045003, 045004.

\bibitem{Hall}G.S. Hall, \textit{Symmetries and Curvature Structure in General Relativity}, World Scientific, Singapore, 2004. 

\bibitem{WMZ}J. Wang, Y. Ma and X.-A. Zhao, Phys. Rev. D $\mathbf{89}$ (2014) 084065.

\bibitem{BF3}J. Wang and C.-G. Huang, Int. J. Mod. Phys. D $\mathbf{25}$, (2016) 1650100.

\bibitem{BF4}C.-G. Huang and J. Wang, Gen. Rel. Grav. $\mathbf{45}$ (2016).

\bibitem{BF5}J. Wang, C.-G. Huang and L. Li, Chin. Phys. C $\mathbf{40}$ (2016).

\bibitem{NP}E. T. Newman and R. Penrose, J. Math. Phys. {\bf 3} (1962) 566.

\bibitem{BCG}R. Basu, A. Chatterjee and A. Ghosh, Class. Quant. Grav. $\mathbf{29}$ (2012), 235010.

\bibitem{HK}Chao-Guang Huang and Shi-Bei Kong, Commun. Theor. Phys. $\mathbf{68}$ (2017), 227.

\bibitem{Dirac}P.A.M. Dirac, \textit{Lectures on Quantum Mechanics}, Yeshiva University, New York, 1964.



\end{thebibliography}
\end{document}